\documentclass[12pt]{article}
\usepackage{cite}

\usepackage{jheppub}
\usepackage{amsmath, amsthm,mathtools,setspace, 
hyperref, url,
amssymb,upgreek,textgreek,
slashed
}

\usepackage[mathscr]{eucal}
\usepackage{tikz-cd}

\parskip=\baselineskip

\begin{document}

\begin{titlepage}
\begin{flushright}

\end{flushright}

\vskip 1 in
\begin{center}
{\bf\Large{A Note On Projective Structures On Compact Surfaces}}

\vskip
0.5cm  { Xiao Liu} \vskip 0.05in {\small{ \textit{School of Physics, \,\,University of Electronic Science and Technology of China}
\vskip .0cm{\textit{Email: $\mathrm{liux@uestc.edu.cn}$}}}}

\end{center}

\vskip 0.3in

\baselineskip 16pt

\begin{abstract}  

Projective structures on topological surfaces support the structure of two-dimensional conformal field theories with a degree of technical simplification.  
By combining a cohomology argument and the uniformization theorem of Riemann surfaces, we define a natural bijective map from the space $\mathcal{P}_g$ of all inequivalent projective structures on the compact topological surface of genus $g$ to the holomorphic cotangent bundle $T^*_{(1,0)} \mathcal{M}_g$ of the moduli space $\mathcal{M}_g$ of Riemann surfaces. Modulo issues at the orbifold loci $\Delta_g$ of $\mathcal{M}_g$ which we analyze, this equips $\mathcal{P}_g$ with the structure of a complex analytic manifold, and qualifies $\mathcal{P}_g$ as a candidate moduli space of projective structures of the genus $g$ topological surface. 

We carry out explicit computations at $g=1$, the only case which also admits affine structures. The affine structure moduli space $\mathcal{A}_{g=1}$ is identified as the global $ {\Bbb Z}_2$ quotient, of a bundle $\Lambda_{\mathcal{M}_{g=1}}$ over $\mathcal{M}_{g=1}$ related to the Hodge bundle. $\Lambda_{\mathcal{M}_{g=1}}$ has the generic fiber ${\Bbb C}$ over the smooth points, and non-generic fibers that are orbifolds of ${\Bbb C}$ over the orbifold loci. Working in local analytic coordinates on $\mathcal{M}_{g=1}$, the projective structure moduli space $\mathcal{P}_{g=1}$ is identified globally with the $T^*_{(1,0)} \mathcal{M}_{g=1}$ including at the fictitious orbifold singularities, and is shown to resolve the $ {\Bbb Z}_2$ orbifold of $\mathcal{A}_{g=1}$. For $g \geq 2$, intricate quotient operations are expected along the fibers over the  orbifold loci $\Delta_g$, with their local analysis left to future work.
 
Physically, the space $\mathcal{P}_g$ represents the bundle of universal, stationary, chiral hydrodynamic flows spatially confined to compact genus-$g$ Riemann surfaces.

   \end{abstract}

\date{October, 2021}
\end{titlepage}

\def\G{{\text{\sf G}}}

\def\la{\langle}
\def\ra{\rangle}
\def\ga{\gamma}
\def\Ga{\Gamma}

\def\veps{\varepsilon}

\def\ad{{\mathrm{ad}}}
\def\SO{{\mathrm{SO}}}
\def\Spinc{{\mathrm{Spin}}_c}
\def\Re{{\mathrm{Re}}}
\def\Im{{\mathrm{Im}}}
\def\SU{{\mathrm{SU}}}
\def\SL{{\mathrm{SL}}}
\def\cl{{\mathrm{cl}}}
\def\d{{\mathrm d}}
\def\dD{{\mathrm D}}
\def\free{{\mathrm{free}}}
\def\sg{{\mathrm g}}
\def\OO{{\mathrm O}}
\def\Sp{{\mathrm{Sp}}}
\def\PSp{{\mathrm{PSp}}}
\def\Spin{{\mathrm{Spin}}}
\def\SL{{\mathrm{SL}}}
\def\SU{{\mathrm{SU}}}
\def\SO{{\mathrm{SO}}}
\def\PGL{{\mathrm{PGL}}}
\def\i{{\mathrm i}}
\def\dim{{\mathrm{dim}}}
\def\spinc{{\mathrm{spin}_c}}
\def\Pic{{\mathrm{Pic}}}
\def\tr{{\mathrm{tr}}}
\def\Pf{{\mathrm{Pf}}}
\def\PSL{{\mathrm{PSL}}}
\def\PSU{{\mathrm{PSU}}}
\def\Im{{\mathrm{Im}}}
\def\Gr{{\mathrm{Gr}}}
\def\sign{{\mathrm{sign}}}
\def\sc{{\mathrm{sc}}}
\def\Max{{\mathrm{Max}}}
\def\Min{{\mathrm{Min}}}
\def\Homeo{{\mathrm{Homeo}}}
\def\CS{{\mathrm{CS}}}
\def\Diff{{\mathrm{Diff}}}
\def\diff{{\mathrm{diff}}}
\def\Sym{{\mathrm{Sym}}}
\def\CS{{\mathrm{CS}}}
\def\CP{{\mathrm{CP}}}
\def\Hom{{\mathrm{Hom}}}
\def\cc{{\mathrm{cc}}}
\def\op{{\mathrm{op}}}
\def\iff{{\mathrm{iff}}}
\def\Arg{{\mathrm{Arg}}}
\def\Tr{{\rm Tr}}
\def\exp{{\rm exp}}
\def\gst{\mathrm{g}_{\mathrm{st}}}
\def\rfa{{\mathrm{for \,\,any}}}
\def\Hom{{\mathrm{Hom}}}
\def\Ext{{\mathrm{Ext}}}
\def\kernel{{\mathrm{kernel}}}
\def\image{{\mathrm{image}}}
\def\Aut{{\mathrm{Aut}}}

\def\fg{{\mathfrak g}}
\def\fsu{{\mathfrak {su}}}
\def\fU{{\mathfrak{U}}}
\def\fh{{\mathfrak{h}}}
\def\fg{{\mathfrak{g}}}
\def\fb{{\mathfrak{b}}}
\def\fc{{\mathfrak{c}}}
\def\fome{{\mathfrak{\omega}}}
\def\fV{{\mathfrak{V}}}
\def\fW{{\mathfrak{W}}}
\def\fk{{\mathfrak{k}}}
\def\fO{{\mathfrak{O}}}
\def\fg{{\mathfrak{g}}}
\def\frak{\mathfrak}
\def\frL{{{\mathfrak L}}}
\def\fD{{\mathfrak D}}
\def\fu{{\mathfrak{u}}}
\def\fe{{\mathfrak{e}}}
\def\ff{{\mathfrak{f}}}
\def\fl{{\mathfrak{l}}}
\def\fp{{\mathfrak{p}}}
\def\hfc{{\hat{\mathfrak{c}}}}
\def\bfc{{\bar{\mathfrak{c}}}}
\def\fh{{\mathfrak{h}}}

\def\cB{{\mathcal B}}
\def\zZ{{\mathcal Z}}
\def\cS{{\mathcal S}}
\def\cD{{\mathcal D}}
\def\cF{{\mathcal F}}
\def\J{{\mathcal J}}
\def\J{{\mathcal J}}
\def\cM{{\mathcal M}}
\def\cT{{\mathcal T}}
\def\V{{\mathcal V}}
\def\E{{\mathcal E}}
\def\N{{\mathcal N}}
\def\B{{\mathcal B}}
\def\cH{{\mathcal H}}
\def\cO{{\mathcal O}}
\def\cP{{\mathcal P}}
\def\Q{{\mathcal Q}}
\def\W{{\mathcal W}}
\def\cM{{\mathcal M}}
\def\cA{{\mathcal A}}
\def\RR{{\mathcal R}}
\def\I{{\mathcal I}}
\def\U{{\mathcal U}}
\def\Bun{{\mathcal M}(G,C)}
\def\L{{\mathcal L}}
\def\O{{\mathcal{O}}}
\def\G{{\mathcal G}}
\def\cL{{\mathcal L}}
\def\tht{{\mathcal \theta}}
\def\cp{{\mathcal p}}

\def\sD{{\mathscr D}}
\def\sO{{\mathscr O}}
\def\sG{{\mathscr G}}
\def\sC{{\mathscr C}}
\def\sM{{\mathscr M}}
\def\sA{{\mathscr A}}
\def\sB{{\mathscr B}}
\def\sE{{\mathscr E}}
\def\sF{{\mathscr F}}
\def\sR{{\mathscr R}}
\def\sPSL{{\mathscr{PSL}}}
\def\sSL{{\mathscr{SL}}}
\def\sGL{{\mathscr{GL}}}
\def\sM{{\mathscr{M}}}

\def\vth{{\vartheta}}

\def\sF{{\sf F}}
\def\sB{{\sf B}}
\def\sA{{\sf A}}
\def\m{{\sf m}}
\def\NS{{\sf{NS}}}
\def\Ra{{\sf{R}}}
\def\sV{{\sf V}}
\def\j{{\sf j}}
\def\sCP{{\sf{CP}}}
\def\S{{\sf S}}
\def\K{{\sf K}}
\def\sfM{{\sf M}}
\def\sfd{{\sf d}}
\def\sP{{\sf P}}
\def\sfT{{\sf T}}

\def\R{{\Bbb R}}
\def\C{{\Bbb C}}
\def\Z{{\Bbb Z}}
\def\RP{{\Bbb{RP}}}
\def\R{{\Bbb{R}}}
\def\bCP{{\Bbb{CP}}}
\def\C{{\mathbb C}}
\def\HH{{\mathbb H}}
\def\Bbb{\mathbb}
\def\3{{\bf 3}}
\def\4{{\bf 4}}
\def\16{{\bf 16}}

\def\1{{(1)}}
\def\2{{(2)}}

\def\h{\widehat}
\def\b{\overline}
\def\u{u}
\def\D{D}
\def\Rf{{\eurm{R}}}
\def\sp{{\sigma}}

\def\bar{\overline}
\def\bg{\bar\ga}
\def\bA{\bar{\mathscr A}}
\def\bB{{\bar {\mathscr B}}}
\def\bop{{\overline{\mathrm{op}}}}
\def\bM{{\overline \M}}
\def\bM{{\overline\M}}
\def\hbbar{\pmb{\hbar}}
\def\dbar{\bar {\partial}}
\def\del{\partial}

\def\til{\tilde}
\def\wtil{\widetilde}
\def\gstt{\widetilde{\mathrm{g}}_{\mathrm{st}}}

\def\hz{{\hat{z}}}
\def\hu{{\hat{u}}}
\def\hh{{\hat{h}}}
\def\hv{{\hat{v}}}
\def\hV{{\hat{V}}}
\def\hT{{\hat{T}}}
\def\tv{{\tilde{v}}}
\def\tV{{\tilde{V}}}
\def\hw{{\hat{w}}}
\def\tg{{\tilde{g}}}
\def\tG{{\tilde{G}}}
\def\th{{\tilde{h}}}
\def\tM{{\tilde{M}}}

\def\PF{{\mathit{P}\negthinspace\mathit{F}}}

\def\be{\begin{equation}}
\def\ee{\end{equation}}

\def\ll{\langle\langle}

\def\rr{\rangle\rangle}
\def\la{\langle}
\def\CP{{C\negthinspace P}}

\def\ra{\rangle}

\def\v{v}

\def\hat{\widehat}

\font\tencmmib=cmmib10 \skewchar\tencmmib='177
\font\sevencmmib=cmmib7 \skewchar\sevencmmib='177
\font\fivecmmib=cmmib5 \skewchar\fivecmmib='177
\newfam\cmmibfam
\textfont\cmmibfam=\tencmmib \scriptfont\cmmibfam=\sevencmmib
\scriptscriptfont\cmmibfam=\fivecmmib
\def\cmmib#1{{\fam\cmmibfam\relax#1}}
\numberwithin{equation}{section}
\def\lmark{{\mathrm L}}

\def\neg{\negthinspace}

\def\be{\begin{equation}}
\def\ee{\end{equation}}

\tableofcontents

\section{Introduction and Summary}\label{intro}

Projective structures on topological surfaces are coordinate coverings of which the transition functions are projective linear. Since projective linear transformations are holomorphic, they support the structure of 2d CFTs \cite{BPZ} and offer a degree of technical simplification by allowing the stress tensor to be analyzed covariantly, on equal footing with non-anomalous primary fields. 

Even though a projective structure is highly special and \emph{non-generic} among the atlases compatible with its underlying holomorphic structure, it is also known to exist on \emph{all} Riemann surfaces (even compact surfaces of $g \geq 2$).\footnote{Where the distinction becomes important, we use $\Sigma$ to denote a topological surface ($\Sigma^{(g)}$ if its genus is important), and $M$ to denote a Riemann surface ($M^{(g)}$ if its genus is important).} . This property of \emph{robust existence} suggests that they have a significant prospect for application, if CFT computations on Riemann surfaces of higher genera are to be pursued.  
Motivated by this, we set out to improve upon the existing understanding of these structures, and have produced, somewhat surprisingly, a quite explicit, geometric description of the full set of these structures.

\vskip 0.7cm

\subsection{$C^1(M, \sO(\kappa^2))$ and Uniformization}\label{firstway}

One way to reveal the robust existence of projective structures on Riemann surfaces is through a standard,  cohomology-type argument. After briefly recalling in section \ref{SD} the basic properties of the Schwarzian operators $S_a$, we put the Schwarzian operator $S_2$ to work in section \ref{PS}. The significance of $S_2$ in this context is that, for the holomorphic structure defined on $M$ by a coordinate covering $\fU=\{(U_i, z_i)\}_{i \in J}$,  the collection of the Schwarzians of the transition functions  $\{\sigma_{2, ij}  = \{z_i, z_j\}_2 \}_{(i,j) \in N(\fU)}$ encodes the obstruction for $\fU$ to be projective.
The ``pseudogroup property'' of $S_2$ (see (\ref{group2})) implies that this collection $\{\sigma_{2, ij}\}_{(i,j) \in N(\fU)}$ in fact forms a cocycle in $Z^1(\fU, \sO(\kappa^2))$, where $\kappa \in H^1(M,\sO^*)$ is the canonical line bundle on $M$, and $\sO^*$ is the sheaf of germs of nowhere vanishing holomorphic functions on $M$. By a combined application of the Serre duality theorem, the Riemann-Roch theorem, and the Chern class formula of a line bundle in terms of its divisors, the cohomology group $H^1(\fU, \sO(\kappa^2))$ is shown to be trivial for all $g \geq 2$. Therefore the obstruction cocycle $\{\sigma_{2, ij}\}$ must be a coboundary, and can thus be absorbed by a set of local changes of coordinate compatible with the given holomorphic structure. A cochain $\{h_i\}_{i \in I} \in C^0(\fU, \sO(\kappa^2))$ that absorbs this coboundary is called a \emph{projective connection}. And the required local holomorphic changes of coordinate are uniquely determined patch-wise, up to projective equivalence, by the projective connection $h$ through the third order differential equation (\ref{u}).

The above reasoning is identical to that given in \cite{Gunning}, which contains excellent information on this and many other topics. A simple by-product of this cohomology argument is the observation that the set of projective connections $h$ forms a complex \emph{linear manifold} (which we call $\cP_{M}^{(0)}$) of complex dimension $\dim H^0(M, \sO(\kappa^2))= 3 g -3$. This is also stated in \cite{Gunning},  without further development down the road.

Based on this, we proceed in section \ref{uniformization} to bring forth a simple geometric description to the set of all projective structures on the genus $g$ compact topological surface $\Sigma^{(g)}$. Our point of departure is the way in which the universal covering space (equipped with varying degrees of structures) of $M^{(g)}$ is put to work. While in section 9(e) of \cite{Gunning} the author applies the universal covering map to pull the projective structure on $M^{(g)}$ back \emph{up} to its universal \emph{topological} cover (which is denoted $\tM$ in \cite{Gunning}) to define on $\tM$ what is called a geometric realization of the projective structure on  $M^{(g)}$, we, on the other hand, apply the profoundly deep and powerful uniformization theorem of Riemann surfaces \cite{FK} \cite{Hubbard} and push the intrinsic projective structure of the universal \emph{analytic} cover \footnote{Which, for $g \geq 2$, is well known to be the open disk $D$ with its standard holomorphic structure inherited from $\C$ by the inclusion $D \subset \C$.} \emph{down} to  $M^{(g)}$. Noting that the covering transformations $\Ga_M$ of the universal analytic covering map $\pi_M: D \to M^{(g)}$ are projective linear, i.e., $\Ga_M \subset \Aut(D) \subset PSL(2, \C)$, this allows us to define a canonical projective structure $\rho^0_{M} \in \cP_{M}^{(0)}$ on $M^{(g)}$,\footnote{This indeed provides an alternative,  more direct way of demonstrating the robust existence of \emph{one} projective structure on each Riemann surface. But it does not have the benefit of determining along the way the full set of \emph{all} possible projective structures on that surface, as does the less direct cohomology analysis.}  and give $\cP_{M}^{(0)}$ the structure of a complex \emph{vector space}.

Given that the difference between two projective connections $h$ is a globally defined holomorphic quadratic differential on $M$, one is led to a natural bijective map from $\cP_g^{(0)\prime}= \bigcup_{[M] \in \cM_g \backslash \Delta_g} \cP_{M}^{(0)}$  to $T^*_{(1,0)} \mathcal{M}_g|_{\cM_g \backslash \Delta_g}$, the holomorphic cotangent bundle of the moduli space of  genus $g$ Riemann surfaces,  with $[M]$ away from the orbifold loci $\Delta_g \subset \cM_g$. Apart from the bundle localized at the orbifold loci for $g \geq 2$, this gives the space $\cP_g^{(0)} = \bigcup_{[M] \in \cM_g } \cP_{M}^{(0)} $ the structure of a complex analytic manifold (of complex dimension $6 g -6$ if $ g \geq 2$, and of complex dimension $2$ if $g=1$). We deem it appropriate to regard $\cP_g$ (an upgraded version of $\cP_g^{(0)}$ by additional quotients, see the next subsection, and subsection \ref{summary}, particularly Item IV) as the moduli space of projective structures on the compact topological surface of genus $g$.
The natural projection map $p:\cP_g \to \cM_g $ sends each projective structure to the holomorphic structure which it is subordinate to.

\vskip 0.7cm

\subsection{Additional Quotients at Orbifold Loci $\Delta_g \subset \cM_g$}\label{orbifold}

The complex structure moduli space $\mathcal{M}_g$ has orbifold singularities for $g \geq 2$, which correspond to Riemann surfaces $M^{(g)}$ admitting nontrivial holomorphic automorphism. The ``holomorphic cotangent bundle'' needs to be properly generalized in the vicinity of these orbifold loci. A parallel complication, but of different interpretation, arises in the study of the projective structures on such symmetric Riemann surfaces, which we now explain. 

Two points on $\cP_M$ correspond to two projective structures that are mutually incompatible on the Riemann surface $M$, because their projective structures are related by a non-vanishing projective connection as in (\ref{u})(\ref{transformed}). But they are not necessarily non-isomorphic. This distinction is standard. \emph{Mutually compatible} means, when we put the two sets of projective atlases together, the additional transition functions one is obligated to introduce to connect the two sets of coordinate maps remain strictly projective. \emph{Isomorphic} (which we use interchangeably with the term \emph{equivalent}), on the other hand, is weaker and means $\Sigma^{(g)}$ admits an automorphism that consists exclusively of projective maps, when its domain is restricted to each coordinate patch and pulled back to $\C$ using the coordinate maps of one structure, and its codomain is restricted accordingly and pushed to $\C$ using the coordinate maps of the other structure\footnote{In the context of affine structures at $g=1$, this distinction explicitly arises in the simple computations of section \ref{torus}. The reader who finds this point confusing is strongly encouraged to visit this example now.}.

Compatible projective structures are automatically isomorphic (the identity map on $\Sigma^{(g)}$ serves as the automorphism), but incompatible projective structures can still be equivalent. The latter possibility arises when nontrivial \emph{projective} automorphisms exist on the surface. This, of course, can happen only for Riemann surfaces admitting nontrivial \emph{holomorphic} automorphisms, i.e. those that correspond to the orbifold loci of $\mathcal{M}_g$ which we collectively denote as $\Delta_g \subset \mathcal{M}_g$. 

Understanding the structure of the projective structure moduli space on such symmetric Riemann surfaces and comparing it with the properly generalized ``holomorphic cotangent bundle'' in the vicinity of the orbifold loci $\Delta_g \subset \mathcal{M}_g$ require detailed local studies that we hope to conduct in future work.  
In absence of these, precise statements can nonetheless be made about what one should expect to happen at the orbifold loci $\Delta_g$, based on general reasoning and on a simple case study at genus 1. However, before more comments are made on this point in subsection \ref{summary}, let us first outline an alternative approach to analyzing the projective structure moduli space, which, inevitably, also leads us to zoom in on the orbifold loci $\Delta_g$ of $\mathcal{M}_g$.

\vskip 0.7cm

\subsection{Coordinate Classes $H^{1 (c)}(M, PSL(2, \C))$}\label{coordinateclasses}

As we will review in Section \ref{ccc}, a natural, intrinsic characterization of the projective structures is provided by the constant-$PSL(2,\C)$-sheaf cohomology set $H^1(\Sigma, PSL(2, \C))$, or equivalently the flat-$PSL(2,\C)$-bundle classes on the topological surface $\Sigma$.\footnote{See the end of section \ref{gstruct} for a clarification of the terminology to avoid any potential confusion.} The projective structures on $\Sigma$ correspond to a special type of classes in $H^1(\Sigma, PSL(2, \C))$, the so-called coordinate classes (collectively denoted as 
 $H^{1 (c)}(\Sigma, PSL(2, \C))$)\footnote{We use $H^{1 (c)}(\Sigma, PSL(2, \C))$ to denote the set of all coordinate cohomology classes on the topological surface $\Sigma$. We use $H^{1 (c)}(M, PSL(2, \C)) \subset H^{1 (c)}(\Sigma, PSL(2, \C))$ to denote the set of coordinate classes on the Riemann surface $M$. These are the classes of projective structures that are compatible with the holomorphic structure on $M$.}.  They are distinguished by having special, global sections (called coordinate sections) that, locally, are homeomorphisms onto their images in $\C$.  
As the discussions (mostly explaining the relevant definitions) in section \ref{gstruct} should make clear,  compatible projective structures must give rise to an identical coordinate class, but the validity of the converse statement is much more subtle.

In fact, non-isomorphic (and therefore automatically incompatible) projective structures can produce an identical coordinate class. According to \cite{Gunning},  it is a direct consequence of the Simultaneous Uniformization Theorem  \cite{Hubbard} \cite{Bers} that Riemann surfaces of the same underlying topology but of distinct holomorphic structures admit projective structures with the same coordinate class in $H^1(\Sigma, PSL(2, \C))$. Therefore $H^{1 (c)}(\Sigma, PSL(2, \C))$ in general is too crude to provide a complete classification of the isomorphism classes of projective structures on the topological surface. 
%far from providing a complete classification of the projective structures, $H^{1 (c)}(\Sigma, PSL(2, \C))$ in general is even too crude to differentiate nonisomorphic structures. 
In particular, the space $\cP_g$ constructed in section \ref{PS} can not be identified with $\bigcup_{[M] \in \cM_g} H^{1 (c)} (M, PSL(2, \C)) \subset H^1 (\Sigma, PSL(2, \C)) $ \footnote{For the meaning of the notation $\bigcup_{[M] \in \cM_g}$ which may be confusing, see footnote \ref{family}.}, because the sets participating this union are not disjoint as the $\cP_M$'s were in the case of $\cP_g$.

If we once again restrict to a single Riemann surface, it turns out that the coordinate classes carry exactly the same information as the compatibility classes. This is a priori too fine for our purpose; we had hoped that they detect only isomorphism, not compatibility. The remedy of the cause on the other hand is apparent. As already explained before, the distinction between compatibility and isomorphism only materializes for surfaces with nontrivial projective automorphisms, and therefore must be localized at the orbifold loci of $\mathcal{M}_g$.

Given that $H^{1 (c)}(M, PSL(2, \C))$  constitutes a very natural construction in itself,  we decide to review and explain two major results about it. The first is a direct algorithm for computing the coordinate cohomology class of a projective structure from the data of its projective connection. This makes explicit the connection between the coordinate class $H^{1 (c)} (M, PSL(2, \C))$ approach and our first $C^1(M, \sO(\kappa^2))$ approach. The second result is the afore-mentioned ``bijection theorem'' that cohomologous structures are necessarily compatible (hence trivially also isomorphic). 

\vskip 0.7cm

\subsection{Summary,  Genus One, and $\dim_\C H^1(\Sigma, PSL(2, \C)) $}\label{summary}

To summarize the general state of affairs:   
\begin{itemize}
\item[(I)]   The space $\cP_M^{(0)}$ is the set of all \emph{incompatible} projective structures on the Riemann surface $M$. It is canonically identified with $H^0(M, \sO(\kappa^2))$, the vector space of holomorphic quadratic differentials on $M$. For two projective structures on the same Riemann surface $M$, the conditions of being compatible and of being cohomologous in $H^{1 (c)}(M, PSL(2, \C))$  are exactly equivalent. 
\item[(II)]   If we use the notation $\cP_M$ to denote the set of all \emph{nonisomorphic} projective structures on the Riemann surface $M$, it can differ from $\cP_M^{(0)}$ \emph{only} for $M$'s that admit nontrivial projective automorphisms. Such surfaces must "reside" at the \emph{orbifold loci} $\Delta_g$ of $\cM_g$.
%then we expect it to be a certain quotient of $\cP_M$, denoted as  $\cP_M^{(0)} =\cP_M/\sim_{\,_M} $.
\item[(III)]   Define $\cP_g = \bigcup_{M \in \tilde{\cM}} \cP_{M}$, where  $\tilde{\cM}$ is a faithful representation of the moduli space $\cM_g$\footnote{By requiring that the family $\tilde{\cM}$ be a faithful representation of the moduli space $\cM_g$, we rule out the possibility of having in the family two Riemann surfaces equipped with isomorphic but incompatible holomorphic structures. We will below also write ``$M \in \tilde{\cM}$'' in a more suggestive but possibly somewhat confusing way as ``$[M] \in \cM_g$'' . \label{family}}. This is the set of all \emph{nonisomorphic} projective structures on the topological surface $\Sigma^{(g)}$.
 
 It agrees with the space $\cP_g^{(0)} = \bigcup_{[M] \in \cM_g} \cP_{M}^{(0)}$ at least for $[M]$ away from the orbifold loci $\Delta_g$ of ${\cM_g}$, where it is canonically identified with $ T^*_{(1,0)} (\mathcal{M}_g \backslash \Delta_g) $.
\item[(IV)]   For a surface $M^{(g)}_s$ admitting nontrivial projective automorphism (therefore necessarily $[M^{(g)}_s] \in \Delta_g$), it is possible that two incompatible projective structures on $M^{(g)}_s$ are nonetheless isomorphic. In this case, a nontrivial quotient is needed to reduce $\cP_{M_s}^{(0)}$ to $\cP_{M_s}$, i.e.  $\cP_{M_s} = \cP_{M_s}^{(0)} / \sim_{M^{(g)}_s}$. Furthermore, the equivalence relation $\sim_{M^{(g)}_s}$  may have an intricate structure as $M^{(g)}_s$ varies on $\Delta_g \subset \cM_g$.
 
\item[(V)]   Understanding the structure of the projective structure moduli space on such symmetric Riemann surfaces and comparing it with the properly generalized ``holomorphic cotangent bundle'' in the vicinity of the orbifold loci $\Delta_g$ of $\mathcal{M}_g$ require detailed \emph{local studies} at $\Delta_g$ that we hope to conduct in future work.  

\end{itemize}

Section \ref{lowg} contains the (mostly elementary) calculations done at $g=1$, which is the only genus that additionally supports affine structures. Even though a simpler structure, the analysis of the moduli space for affine structures serves to illustrate the general phenomena concerning the existence of isomorphisms between incompatible structures, and the need for additional quotients at the orbifold loci in the complex structure moduli space $\Delta_{g=1} \subset \cM_{g=1}$. In fact, the affine structure moduli space $\cA_{g=1}$ has a new feature: an extra, \emph{global} $\Z_2$ quotient must be performed over all of $\cM_{g=1}$, a result of the existence of the $\Z_2$ ``parity'' automorphism (i.e. $z \to -z$ in the uniformizing coordinate) on \emph{all} $g=1$ surfaces. 

More explicitly, before taking this global $\Z_2$ quotient, we have a generic fiber $\Ga(\sfT^2_\tau, \sO(\kappa)) = \Ga(\sfT^2_\tau, \sO)= \C$ worth of inequivalent affine structures at each smooth point of $\cM_{g=1}$. As is well-known, the orbifold loci  $\Delta_{g=1} \subset \cM_{g=1}$ in this case consists of two points, one of order 2 (which we call $A$), the other of order 3 (which we call $B$). As explained above\footnote{Albeit in the context of projective structures. The underlying idea clearly also applies to the case of affine structure. \label{affinevsproj}}, we expect additional isomorphisms of affine structures at $\Delta_{g=1}=\{A, B\}$, and indeed simple analysis shows that the fiber at $A$ is reduced to $\C/\Z_4$, while at $B$ it is reduced to $\C/\Z_6$. The resulting bundle (having a generic fiber $\C$ at each point of $\cM_{g=1}\backslash \Delta_{g=1}$, a $\C/\Z_4$ fiber at $A$, and a $\C/\Z_6$ fiber at $B$), which we denote as $\Lambda_{\cM_{g=1}}$, appears to be related to the Hodge bundle in the math literature (see for example \cite{HM}). The global $\Z_2$ quotient then reduces the generic fibers to $\C/\Z_2$, while leaving untouched the non-generic fibers at  $\Delta_{g=1}=\{A, B\}$, producing a bundle we denote as $\Lambda_{\cM_{g=1}}/\Z_2$.

The projective structure moduli space $\cP_{g=1}$ is then analyzed. It turns out that $\cP_{g=1}$ provides a resolution of the global $\Z_2$ orbifold action in such a way that it produces a bundle whose generic fiber is $\Ga(\sfT^2_\tau, \sO(\kappa^2)) = \Ga(\sfT^2_\tau, \sO)= \C$, and whose non-generic fibers are $\C/\Z_2$ and $\C/\Z_3$ at $A$ and $B$ respectively. When switching to local analytic coordinates near the orbifold loci so that $\cM_{g=1}$ (including the points $A$ and $B$) becomes biholomorphic to $\C$, one sees that the total projective structure moduli space $\cP_{g=1}$ agrees exactly with the analytic space $T^*_{(1,0)} \cM_{g=1}$.

As we go to higher values of the genus,  explicit computations still seem feasible if $g$ is moderate. But for significantly larger values of $g$, more power technologies seem indispensable for the analysis of the orbifold loci $\Delta_g \subset \cM_g$. One would also like to study the projective structures on degenerate Riemann surfaces, which correspond to measure-zero boundaries of properly compactified moduli spaces $\bar{\cM}_g$ (or $\bar{\cM}_{g,n}$ if with punctures) \cite{HM} \cite{FriedanShenker}.  This also requires more advanced algebraic geometry technologies. 

Short of these powerful tools, we switch track 
and work out some simple semi-quantitative result in the final section \ref{highg}. A natural question one may ask is: how special are the sets $H^{1 (c)}(M, PSL(2, \C))$ and $H^{1 (c)}(\Sigma, PSL(2, \C))$ of coordinate classes as subspaces of $H^1(\Sigma, PSL(2, \C))$? A crude answer is provided by the dimensionality of the respective spaces. The bijection relation of section \ref{uniq} equips $H^{1 (c)}(M, PSL(2, \C))$ with the structure of a complex analytic manifold of complex dimension $3 g -3$. A simple calculation in section \ref{highg} shows that  $H^1(\Sigma, PSL(2, \C))$ as a space naturally has complex dimension $6 g -6$. So under the assumption that the structure identified on $H^{1 (c)}(M, PSL(2, \C))$ from its relation to $H^0(M, \sO(\kappa^2))$ is the same as the structure induced from the inclusion map, we conclude that $H^{1 (c)}(M, PSL(2, \C))$ is a middle-dimensional subspace of $H^1(\Sigma, PSL(2, \C))$.

In the appendixes, we briefly review and explain some of the mathematical facts that we have made essential use of in the main text.

\vskip 0.6cm

\subsection{A Second, ``\emph{Dual}'', Physics Application}

The above summarizes the mostly mathematical discussions of the projective structures contained in the main text, with its most direct application to 2d CFTs in mind. Next, however, we would like to present an alternative, \emph{``dual''} perspective on the connection between the projective structures and the physics of two dimensional CFTs.

The (holomorphic) energy-momentum tensor operator $\hT_{zz}$ of a 2d CFT on a Riemann surface $M^{(g)}$ \emph{is} an \emph{operator-valued projective connection} in itself. As is well-known, while in each chart of the holomorphic structure it is represented by a holomorphic quadratic differential
\be\label{quadraticdiff}
\hT_{zz}(z) \d z^2,
\ee 
its values in the intersection of two adjacent charts are related by 
\be\label{anomal}
\hT_{uu}(u) \d u^2 =  \hT_{zz}(z) \d z^2 - \frac{c}{12} \{u,z \} \d z^2,
\ee
where $\{u,z \}$ is the Schwarzian derivative (\ref{schwarzian}) (\ref{schwarzian2}). If one compares (\ref{anomal}) and (\ref{connection}), the only major difference is that  $\hT$ is a local holomorphic operator in $U$ while $h$ a local holomorphic function.

This suggests a second, \emph{dual} application of the space $\cP_g$ of projective connections. In stead of the above, standard approach of using the projective connections to generate the \emph{non-generic, special} projective coordinate systems in which the Schwarzian derivative terms vanish in (\ref{anomal}), one can use them as the space of classical background solutions of $\hT_{zz}$. Namely we set\footnote{Here notation-wise, we write $h$ with two lower indices. We do not have to do so in section \ref{sigma} because there we define $\{h_i\}_{i \in I}$ to be the local sections of (the sheaf of germs of holomorphic sections of) the line bundle $\kappa^2$.}
\be\label{fluc}
\hT_{zz}(z) = \frac{c}{12} \cdot  h_{zz}(z) + \hT_{zz}^q (z)
\ee
As $h_{zz}(z)$, being a projective connection, satisfies (\ref{connection}), or more explicitly 
\be\label{anomal2}
h_{uu}(u) \d u^2 =  h_{zz}(z) \d z^2 - \{u,z \} \d z^2,
\ee
the ``quantum'' component $\hT_{zz}^q (z)$ of the energy-momentum tensor reduces to a genuine globally defined operator-valued holomorphic quadratic differential in \emph{all} holomorphic coordinate systems
\be\label{2form}
\hT_{uu}^q (u) \d u^2 =  \hT_{zz}^q (z) \d z^2.
\ee
The background solutions 
\be\label{background}
T^c_{zz}(z) =\frac{c}{12} \cdot  h_{zz}(z)
\ee
may be interpreted as the one-point expectation value of the operator $\hT_{zz}(z)$ produced, for example, by a Euclidean path integral of the CFT on the Riemann surface $M^{(g)}$. And it is expected to describe the classical dynamics of the CFT in the ``hydrodynamic'' regime \footnote{We put the quotation marks because this ``hydrodynamics'' resides in a 2d Euclidean space without time. To make contact with real hydrodynamics, one needs to consider $2+1$ dimensional fluid systems in the state of stationary flows.}.

Given the ubiquity of the energy-momentum tensor operator as a universal sector in the operator algebra of all two dimensional CFTs, the space $\cP_g$ of projective connections that we constructed, up to the  normalization factor $\frac{c}{12}$, truly represents the bundle of universal chiral ``hydrodynamic'' solutions of all two-dimensional conformal field theories on compact genus-$g$ Riemann surfaces.  Connections to real-world physics systems are to be sought in effective $2+1$ dimensional flow systems that are quasi-stationary in time, confined within a two-dimensional space with nontrivial topology, and demonstrating approximate scale invariance. We leave this search to future work. Of course, we expect dissipative effects become important and produce corrections to the solutions as we go far enough out in the fiber directions.

\vskip 0.6cm

\emph{Note on Terminology} \hskip 0.3cm
To avoid any potential confusion, we recall that the construction of projective bundles of various ranks and their associated connections have been discussed very extensively in the classic literature on 2d conformal field theories. In these works \footnote{See for example \cite{FriedanShenker} \cite{Vafa} \cite{MooreSeiberg1} \cite{Segal} \cite{Witten} and  \cite{Verlinde}, from a golden age of the subject.  The classic work \cite{Segal},  also made a passing mention of the projective structures on the Riemann surface itself.}, the projective objects of concern are defined over the moduli space $\cM_g$ of Riemann surfaces, and characterize profound intrinsic properties of 2d conformal theories.  In this note however, they are only confined to a single Riemann surface $M^{(g)}$.

\section{The Schwarzian Derivative $S_2$}\label{SD}

If $f$ is a holomorphic function with $f' \neq 0$ in some region of the complex plane $\C$, one can define the so-called Schwarzian derivative:
\be\label{schwarzian}
(S_2 f) (z) = \{f, z\}_{_2} = \frac{f''' (z) }{f'(z)} -\frac{3}{2} \left(\frac{f''(z)}{f'(z)} \right)^2 . 
%= \frac{2 f^{(1)}(z)  f^{(3)} (z) - 3 f^{(2)}(z)^2}{2 f^{(1)}(z)^2}= \frac{f^{(3)} (z) }{f^{(1)}(z)} -\frac{3}{2} \left(\frac{f^{(2)}(z)}{f^{(1)}} \right)^2 .
\ee
We have introduced two notations for the Schwarzian derivative, and will use whichever that is the more convenient in a given context. 
 
$S_2 f$ has a quite remarkable ``pseudo-group'' property. Let $h=g \circ f$, which we also denote by $z \xrightarrow{f} u \xrightarrow{g} v$.  One can check by direct computation that 
\be\label{group2}
\{v, z \}_{_2} = \{v, u\}_{_2} (u'_z)^2  + \{u,z\}_{_2}.\ee

This relation underlies the consistency of the anomalous transformation of the stress tensor $T(z)$ \footnote{We suppress the tensorial indices of $T$ and use the local holomorphic coordinate variable to simultaneously indicate both the position of the operator and the local frame in which it is evaluated. A more complete notation for the situation in (\ref{stress}) 
is $T_{uu} (u)$ vs. $T_{zz}(z)$.} of a $2d$ CFT:
\be\label{stress}
(\del_z u)^2 T(u) = T(z)-\frac{c}{12} \{u,z\}.
\ee 
Indeed, under $z \xrightarrow{f} u \xrightarrow{g} v$,
\begin{align}\label{consist} \nonumber
T(z) - \frac{c}{12} \{v,z\}  = & (v'_z)^2 T(v)  = (u'_z)^2 (v'_u)^2  T(v) \\ \nonumber
                             = & (u'_z)^2 (T(u)-\frac{c}{12} \{v,u\}) \\  % \nonumber
                             = &  T(z) - \frac{c}{12} \{u,z\} - \frac{c}{12} (u'_z)^2 \{v,u\},
\end{align}
which is consistent by virtue of (\ref{group2}).

$S_2 f$ admits a simpler expression. Since we have to assume in the first place that  $f'(z) \neq 0$ in the domain of interest, we can choose a branch and take its square-root. Then 
\be\label{schwarzian2}
(S_2 f )(z) = -2 (f'(z))^{1/2} \frac{\d^2}{\d z^2} (f'(z))^{-1/2}.
\ee
This makes it easier to solve differential equations of the type 
\be\label{ODE}
(S_2 f) (z) = h(z)
\ee
which is important to us below. For now, we use (\ref{schwarzian2}) to read off the general solution to $$ (S_2 f) (z) =0.$$  
We deduce $f'(z) = (c z+d)^{-2}$  and hence  
\be\label{PL}
f(z)= \frac{a z+ b}{c z + d},\,\,\,\,\,\,\,\,\,\,\,\,\,\,a d - b c =1.\ee
It contains three integration constants, consistent with the fact that  $S_2$ is a third order differential operator. We are starting to see a connection between the Schwarzian derivative $S_2$ and the projective transformations.

\subsection{The $S_1$ Operator}\label{S1}

A minor digression is perhaps in order. A simpler operator  
\be\label{S1}
(S_1 f) (z) = \{f, z \}_{_1} =  \frac{f''(z)}{f'(z)}\ee
shares a ``pseudo-group'' property similar to that of (\ref{group}): under $z \xrightarrow{f} u \xrightarrow{g} v$
\be\label{group1}
\{v, z \}_{_1} = \{v, u\}_{_1} (u'_z)  + \{u,z\}_{_1}.\ee 
For this reason, we adopt the obvious notation $S_a$ and $\{, \}_{_a}$ for $a=1,2$ to denote these two operators, and now the ``pseudo-group'' relation reads
\be\label{group}
\{v, z \}_{_a} = \{v, u\}_{_a} (u'_z)^a  + \{u,z\}_{_a}, \,\,\,\,\,\,\,\,\,\,a=1,2.
\ee 
The solutions to $(S_1 f) (z) =0$ of course are the affine transformations 
$$ f(z) = a z +b,\,\,\,\,\,\,\,\,\,\,\,\,\,\,\,\, a \neq 0.$$

\section{Projective Structures on Compact Riemann Surfaces}\label{PS}

Now we consider a compact Riemann surface $M$. By definition, it is a two dimensional (connected, Hausdorff, compact) topological surface with an atlas of coordinate maps  $\fU=\{U_i, z_i\}_{i \in I}$. 
The $U_i$'s are open sets that, together, cover $M$. $z_i$ is a homeomorphism from $U_i$ to an open set $z_i(U_i) \subset \C$. And the transition functions 
$\{ f_{ij} \triangleq z_i \circ z_j^{-1}: z_j(U_i \cap U_j) \to z_j(U_i \cap U_j)\}_{(i,j) \in N(\fU)}$ are required to be holomorphic in their respective domains of definition\footnote{The complete definition also needs to maximize the atlas so that it contains all compatible coordinate systems. This means that we need to include all the charts that have holomorphic transition functions among themselves. We in general will omit such statements in our discussion, unless we think an omission may cause confusion.}. Put simply, a Riemann surface is a topological surface with a choice of complex analytic structure.  

The claim is that, given such a surface, there always exists further refined choices of coordinate maps $\fU=\{U_i, z_i\}_{i \in I}$ that 
\begin{itemize}
\item[(i)]  are compatible with the given complex structure, and 
\item[(ii)] have transition functions that are exclusively projective linear, i.e. each $f_{ij} \triangleq z_i \circ z_j^{-1}$  takes the form of (\ref{PL}) on $z_j(U_i \cap U_j) \subset \C$.
\end{itemize}
In other words, there always exists a projective structure \emph{subordinate} to the complex structure of $M$. In fact, for each $g \geq 1$,  a continuum infinity of mutually incompatible projective structures exist on each $M$.  

The argument is a beautiful application of a number of fundamental results of Riemann surfaces, and of the machinery of sheaf cohomology, some of which are briefly reviewed and summarized in the Appendixes. We mostly follow the beautiful exposition of \cite{Gunning}.

\subsection{Existence of Projective Structures, Coboundary $\sigma_{aij}$ of $\sO(\kappa^2)$,\\
            and  Projective Connections}\label{sigma}

Let $\kappa$ be the canonical line bundle, equivalently, the line bundle of holomorphic 1-forms. Given a sufficiently fine \footnote{Technically, a Leray covering of $\sO^*$. In practice, we will choose a finite open covering of which all nonempty intersections are contractible.}holomorphic coordinate covering $\fU=\{U_i, z_i\}_{i \in I}$, $\kappa$ is defined by the cocycle $\{\kappa_{ij}= f'_{ij}(z_j(p))^{-1}, p\in U_i \cap U_j \neq \emptyset\} \in Z^1(\fU, \sO^*)$, where $\sO^*$ is the multiplicative sheaf of germs of non-vanishing holomorphic functions. 

Consider %$
\be\label{canonical}\sigma_{a\,ij} (z_j) \triangleq (S_a f_{ij}) (z_j) \hskip 1cm a=1,2 \ee
%$ ($a=1,2$ as before) 
defined in $z_j (U_i \cap U_j)$. The ``pseudogroup'' property (\ref{group}) implies that, when considered as sections of the line bundle $\kappa^a$, they form a 1-cocycle of $Z^1(\fU, \sO(\kappa^a))$. This means \footnote{When affirming that (\ref{cocycle}) indeed follows from (\ref{group}), sufficient attention needs to be paid to the order of the indices on the restriction homomorphisms ($\kappa^a$ in this equation), and the coordinate variables $z_i$ carried both by the restriction homomorphisms and by the sections, which, practically, may be considered as a part of the label on the sets \{$U_i$\}. The same comment applies when checking similar relations later.}
\be\label{cocycle}
\sigma_{a\,ik} (z_k(p)) =\kappa_{kj} (z_j(p))^a \cdot \sigma_{a\,ij} (z_j(p))  +  \sigma_{a\,jk} (z_k(p)), \hskip 1.2cm  \mathrm{for} \,\,\,z_k \in U_i \cap U_j \cap U_k \neq \emptyset.  
\ee

The Serre duality (see Appendix \ref{SDT}) implies that for any line bundle $\xi \in H^1(M, \sO^*)$,
\be\label{serre}
\dim_\C H^1(M, \sO(\xi)) =  \dim_\C H^0(M, \sO(\kappa \xi^{-1})).
\ee
Setting $\xi =\kappa^a, a=1,2$ it gives 
\be\label{h1}
\dim_\C H^1(M, \sO(\kappa^a)) = \dim_\C H^0 (M, \sO(\kappa^{1-a}))\ee
From the Riemann-Roch theorem (see Appendix \ref{RRT}), we know $c(\kappa) = 2 g -2$, hence $c(\kappa^{1-a})=2 (1-a) (g-1)$. 
For $a=2$ and $g \geq 2$, we have $c(\kappa^{-1}) <0$. $\kappa^{-1}$ therefore can not have any nontrivial holomorphic section.\footnote{Recall that the Chern class of a holomorphic line bundle $\xi \in H^1(M, \sO^*)$ can be computed by taking any nontrivial meromorphic section $f \in \Ga(M, \sM^*(\xi))$ and summing up the degrees of all of its divisors: 
\be\label{chern}
c(\xi)=\sum_{p \in M} \nu_p(f).\ee} 
Hence $\{0\} = H^0(M, \sO(\kappa^{-1})) = H^1(M, \sO(\kappa^2))$.  Therefore the cocycle 
$\{\sigma_{2\,ij} (z_j) \triangleq (S_2 f_{ij}) (z_j)\}_{(i,j) \in N(\fU)}$ must be trivial, i.e. being a coboundary in $B^1(\fU, \sO(\kappa^2))$.
%\footnote{For the $g =0, 1$ cases, the cocycle $\sigma_2$ is clearly trivial. Even though the present, general argument does not apply, the result is obvious if it is computed in the standard coordinates and directly from the definition (\ref{canonical}).}
\footnote{For $g=0$ and $1$, $H^0(M, \sO(\kappa^{-1}))=\C^3$ and $\C^1$ respectively. So the general argument given here does not apply. But these two spaces are clearly projective in their respective standard coordinate. And this is easily seen to be consistent as the cocycle $\sigma_2$, readily computed from definition, vanishes in their respective standard coordinate.}

This means that there must exist $\{h_i (z_i)\}_{i \in I} \in C^0(\fU, \sO(\kappa^2))$ such that 
\be\label{connection0}
(\delta h)_{ij}=\sigma_{2\,ij}. \ee
Written out explicitly, this means 
\be\label{connection}
\sigma_{2\,ij}(z_j) = - \kappa_{ji} (z_i)^2 \cdot h_i(z_i) + h_j(z_j) = - f'_{ij} (z_j)^2 \cdot h_i(z_i) + h_j(z_j) .
\ee
The $\{h_i (z_i)\}_{i \in I} \in C^0(\fU, \sO(\kappa^2))$ with this property is called a projective connection. 

Let us imagine we have solved (\ref{ODE}) in each patch $U_i$ and have found $u_i$ satisfying
\be\label{u}
(S_2 u_i) (z_i) = h_i(z_i)
\ee
We claim that $\{ u_i \circ z_i (p), p\in U_i\}_{i \in I} $ is a good projective coordinate atlas. 

This is easy to verify. The new transition functions $\tilde{f}_{ij}$ satisfy
\be\label{transformed}
\tilde{f}_{ij} \circ u_j = u_i \circ f_{ij}
\ee 
Taking the Schwarzian derivative with respect to $z_j$ on both sides and applying the ``pseudogroup'' property to both, we arrive at
\be\label{c}
\{u_i, u_j \}_{_2} (\del u_j / \del z_j)^2 + \{u_j, z_j\}_{_2} = \{u_i, z_i\}_{_2} (\del z_i / \del z_j)^2  + \{z_i,z_j\}_{_2},\ee
which is just the statement
\be\label{c2}
\{u_i, u_j \}_{_2} (\del u_j / \del z_j)^2 = \sigma_{2\,ij} - (h_j(z_j)-  \kappa_{ji} (z_i)^2 \cdot h_i(z_i)) =0.
\ee Hence, assuming (\ref{u}) can be solved,  $\{\tilde{f}_{ij} = ( u_i \circ z_i ) \circ (u_j \circ z_j)^{-1},  (i,j) \in N(\fU)\}$ are indeed projective linear transformations (recall the discussion around (\ref{PL})). 

Making use of (\ref{schwarzian2}), introduce $v_i = (\del u_i/ \del z_i)^{-1/2}$. We need $u_i: z_i(U_i) \to u_i \circ z_i (U_i)$ to be a holomorphic homeomorphism, so the derivative can not vanish in its domain of definition. Assuming this condition is satisfied, which will be checked, the square-root is taken by simply choosing a branch on each $U_i$.  Then (\ref{u}) becomes
\be\label{v}
2 v''_i(z_i) +  h_i(z_i)\cdot v_i (z_i) =0. 
\ee
By further letting $w_i(z_i) = \left[ \begin{array} {c}  v_i(z_i) \\  v_i'(z_i) \end{array} \right]$,
we arrive at 
\be\label{w}
\del  w_i(z_i)/ \del z_i =  H_i(z_i) w_i (z_i),
\ee with
\be\label{H}
H(z_i)= \left[ \begin{array} {cc}  0 & 1 \\  -\frac{1}{2}h_i(z_i) & 0 \end{array} \right].
\ee
(\ref{w}) can be explicitly integratged using path-ordered line integral. Combined with its  $\del /\del \bar{z}$ counterpart, which has a zero connection and just imposes that $w_i$ be holomorphic, it describes the transport of a 2-vector $w_i$ by the \emph{flat} connection $(H_i, \bar{H}_i(=0))$. As long as we choose $U_i$ contractible and small enough, the analyticity of (\ref{H}) clearly implies that $v_i$ is contained in a small enough domain so that $\del u_i/ \del z_i$ never reaches zero and that the square-root in $v_i = (\del u_i/ \del z_i)^{-1/2}$ is well-defined.

Once we have $v_i(z_i)$, we can simply invert $v_i = (\del u_i/ \del z_i)^{-1/2}$ to obtain
\be\label{u2}
u_i(z_i) = \int^{z_i}_{p_0} \d z /v_i^2.
\ee 
The holomorphicity of $v_i$ together with the contractibility of $U_i$ means that the choice of path for the integral has no effect on the result.

This completes the proof that projective structures subordinate to the complex structure on any given compact Riemann surface of $g \geq 2$ always exists.

\subsection{The Space of Projective Structures, Uniformization, and $T^*_{(1,0)} \mathcal{M}_g$}\label{uniformization}

Having derived one solution $u_i(z_i)$ to (\ref{u}), the space of all solutions is easily determined. Any other solution $\tilde{u}_i$, which also has to be a holomorphic homeomorphism, is necessarily of the form 
\be\label{g}
\tilde{u}_i(z_i) = (g \circ u_i) (z_i), 
\ee  
with $g$ a holomorphic homeomorphism. Applying the Schwarzian derivative to both sides of this equation, making use of (\ref{u}) and the ``pseudo-group'' property (\ref{group}), we get
\be\label{g2} 
h_i(z_i) = \{\tilde{u}_i, z_i\}_{_2} = \{\tilde{u}_i, u_i\}_{_2} \cdot (\del u_i/ \del z_i)^2 + \{u_i, z_i\}_{_2} = (S_2 g) (u_i) \cdot (\del u_i/ \del z_i)^2 + h_i(z_i ).
\ee 
Hence $(S_2 g) (u_i) =0$, $g$ must be a projective linear transformation. Different solutions to (\ref{u}) thus define the same projective structure on $M$.

The lesson therefore is: there is a one-one correspondence between the projective connections $\{h_i (z_i)\}_{i \in I} \in C^0(\fU, \sO(\kappa^2))$ satisfying  $(\delta h)_{ij}=\sigma_{2\,ij}$,
and the projective structures subordinate to $M$. The explicit projective coordinate mappings are constructed by solving (\ref{u}) on each patch. These statements hold for all $g \geq 0$.

\vskip 1cm

The projective connections $\{h_i (z_i)\}_{i \in I} \in C^0(\fU, \sO(\kappa^2))$ are only constrained to satisfy $(\delta h)_{ij}=\sigma_{2\,ij}$. Given one solution $h_0$ to (\ref{connection0}), one can construct all other solutions by simply adding an arbitrary global section  $s \in \Ga(\fU, \sO(\kappa^2))=H^0(\fU, \sO(\kappa^2)) $  of (the sheaf $\sO(\kappa^2)$ of germs of holomorphic sections of) the holomorphic line bundle $\kappa^2$. These are precisely the holomorphic quadratic differentials defined globally on $M$. The dimension of the complex vector space they span is again computed by the Riemann-Roch theorem, and is given by $D_g=3 g -3$ for $g \geq 2$, $D_1= 1$ for $g=1$,\footnote{$D_1=1$ because for $g=1$, in addition to $c(\kappa)=0$, we have $\kappa = 1 \in H^1(\sfT^2_\tau, \sO^*)$, i.e. the canonical line bundle in this case is the trivial holomorphic line bundle. This is seen to be the case by noting that there exists a nowhere vanishing holomorphic 1-form on $\sfT^2_\tau$. The generic $c=0$ line bundles on $\sfT^2_\tau$, on the other hand, have no nontrivial global section. \label{torusline}} and $D_0=0$ for $g=0$. Therefore, the space $\cH_M$ of all projective connections $h$ on $M$ is an \emph{affine space} of positive dimensionality $D_g$ if $g \geq 1$, and it is a single point if $g=0$.

Without identifying a canonical projective connection on $M$, this is the best that can be done \cite{Gunning}\footnote{And to my knowledge, this is also the best that has been explicitly stated in the literature. }. Here we propose to apply the powerful uniformization theorem of Riemann surfaces \cite{FK, Hubbard} to define a distinguished projective structure on $M$. 

Specifically, any compact surface of $g \geq 2$ has the open unit disk $D \subset \C$ as its universal analytic covering space. The universal holomorphic covering map $\pi: D \to M$ is a local isomorphism that effectively identifies $M$ with $D / \Ga$, where $\Ga$ is the group of covering transformations. It is a torsion-free, discrete subgroup of $ \Aut(D) \subset PSL(2, \C)$ (i.e. a torsion-free Fuchsian group) isomorphic to the fundamental group of $M$. By definition, each point $p \in M$ has an open neighborhood $U_p$ that is evenly covered by $\pi$; its inverse image $\pi^{-1}(U_p)$ is the union of disjoint open sets $V_{p,\alpha}$, and for each $\alpha$, the restriction $\pi|_{V_{p,\alpha}}$ is a holomorphic homeomorphism of $V_{p,\alpha}$ onto $U_p$. Because the covering transformations are realized by projective linear transformations (i.e. $\Ga  \subset \Aut(D) \subset PSL(2, \C)$  ),  the atlas $\{U_p,\,\,\, (\pi|_{V_{p,\alpha}})^{-1}: U_p \to V_{p,\alpha} \}_{p \in M}$ is a projective coordinate covering of $M$, and hence it defines a projective structure of $M$. 

The complex structure of $M$ is encoded in the covering map $\pi$, or more directly, in the conjugacy class of $\Ga$ inside $\Aut D$. The canonical projective structure is subordinate to the complex structure of $M$, because  
it is defined by locally inverting the covering map $\pi$. To summarize, we have identified a canonical projective structure on $M$ subordinate to its complex structure.\footnote{A similar approach clearly applies to $g=1$, with the universal analytic covering space being $\C$, and $\Aut \C \subset PSL(2, \C) $ consisting of the affine transformations. The group $\Ga$ of covering transformations is isomorphic to $\Z \oplus \Z$, and its conjugacy class in $\Aut \C$ determines the complex structure modulus $\tau$. } 

This simple construction allows us to identify a \emph{canonical} projective connection $h_{\ast,M} \in C^0(\fU, \sO(\kappa^2))$ that corresponds to the canonical projective structure.\footnote{Here to fully appreciate the canonical nature of $h_{\ast,M}$, it is helpful to choose $\fU$ to be the \emph{maximal} atlas $\hat{\fU}$ compatible with the given holomorphic structure on $M$. } 
The space $\cH_M$ of projective connections is now identified canonically with the complex vector space $H^0(M, \sO(\kappa^2))$ of dimension $D_g$ ($3 g -3$ if $g \geq 2$, and $1$ if $g=1$). We will henceforth denote it as $\cH_{M,\ast}$. 

The bijective map between this vector space and the set of projective structures allows us to topologize the latter set, producing a topological space $\cP_{M}^{(0)}$ so that the bijective map becomes a homeomorphism.
On the other hand, given the nonlinear nature of (\ref{ODE}), the natural operation on the space of projective structures (composition of maps) does \emph{not} seem to be related to the natural vector space operations (addition and scalar multiplication) on $\cH_{M, \ast}$ in any simple way. We have therefore yet to see the value of additionally transporting the vector space structure of $\cH_{M,\ast}$ to $\cP_{M}^{(0)}$.

As we vary the surface $M^{(g)}$ over a family $\tilde{\cM}$ of Riemann surfaces that forms a faithful representation of the moduli space $\cM_g$,\footnote{The reason for choosing the family $\tilde{\cM}$ in this way is explained in the footnote \ref{family}.} the $\cH_{M, \ast}$'s pull back to a vector bundle $\cH_\ast$ over the moduli space $\cM_g$. We identify this bundle $\cH_\ast$ with the holomorphic cotangent bundle $T^*_{(1,0)} \mathcal{M}_g$ over $\cM_g$, at least when $[M] \in \mathcal{M}_g$ is away from the orbifold loci $\Delta_g \subset \cM_g$ where the definition of $T^*_{(1,0)} \mathcal{M}_g|_{\Delta_g}$ needs extra care. Recall that the tangent space to $\cM_g$  at $M \in \tilde{\cM}$ consists of the classes of the Beltrami-differentials $[\mu] \in H^1(M, \sO^{-1,0}) = \Ga(M, \sE^{-1,1})/\dbar \Ga(M, \sE^{-1,0})$, which, by Serre's duality, are canonically dual to the holomorphic quadratic differentials $\Ga(M, \sO^{(2,0)})= \Ga(M, \sO(\kappa^2))$:
\be\label{dual}
<\mu, \phi> = \int_M \mu^z_{\bar{z}}  \phi_{zz} \d z \wedge \d \bar{z}. 
\ee
We stress that, by definition, the identification between $\cH_\ast$ and $T^*_{(1,0)} \mathcal{M}_g$ is a genuine isomorphism between two vector bundles on $\cM_g \backslash \Delta_g$. 

By combining with the natural homeomorphism between $\cH_{M, \ast}$ and $\cP_M^{(0)}$ of each $M$, we now have a natural bijective map from the space of all incompatible projective structures 
on the family $\tilde{\cM}'$ of Riemann surfaces (for now excluding from $\tilde{\cM}$ surfaces with extra discrete holomorphic automorphisms, hence the extra notation $'$)
to the holomorphic cotangent bundle $T^*_{(1,0)} \mathcal{M}_g$ over the smooth part of the moduli space:
\be\label{projec}
\Phi: \cP_g^{(0)} \triangleq \bigcup_{M \in \tilde{\cM}'} \cP_{[M]}^{(0)} \to T^*_{(1,0)} \mathcal{M}_g|_{\cM_g \backslash \Delta_g}
\ee

Further exploiting this bijection, we can define the topology and the complex analytic structure on $\cP_g^{(0)}$ in such a way that $\Phi$ becomes a homeomorphism and then an analytic isomorphism. This equips $\cP_g^{(0)}$ with the structure of a complex $2 D_g$ dimensional complex analytic manifold. We think it is appropriate to call $\cP_g^{(0)}$ a \emph{pseudo} moduli space of projective structures of genus $g$.

\section{The Coordinate Cohomology Classes}\label{ccc}

It is natural to expect that the projective structures on a \emph{topological} surface $\Sigma$ can be described by a particular subset of classes in the set $H^1(\Sigma, PSL(2, \C))$, which will be called the coordinate cohomology classes (or the coordinate classes for short). 

Here inside the bracket of $H^1(\Sigma, \bullet)$,  $PSL(2, \C)$ stands for the constant sheaf of the group. $H^1(\Sigma, PSL(2, \C))$ is the cohomology \emph{set} of this sheaf; the cohomology is not a group because $PSL(2, \C)$ is non-abelian.  The constant sheaf, as a set, is the Cartesian product $\Sigma \times PSL(2, \C)$. As a space, its topology is the product topology of the standard topology on $\Sigma$ and the \emph{discrete} topology on $PSL(2, \C)$. This forces any section over a connected open set to be a constant function, taking the value of a particular group element. This is the reason for the ``degeneracy'' of the notation.

\subsection{$H^1(\Sigma, G)$, G-Structure, and Flat $G$-Bundles }\label{gstruct}

The definition of $H^1(\Sigma, G)$ for a nonabelian group $G$ is standard. We briefly sketch it here, pointing out the differences from the case of constant sheaves of abelian groups. It involves two steps as usual. First, one computes the cohomology sets $H^1(\fU, G)$ for \emph{all} open covers $\{ \fU \}$ of $\Sigma$. Second, noting that under an arbitrary refinement $\fU < \fV $ (we use this to indicate that $\fV $ refines $\fU$), the natural map $\mu^*: H^1(\fU, G) \to H^1(\fV, G) $ is independent of the choice of the refining map, one takes the direct limit
\begin{align}\label{directlimit}\nonumber
&\varinjlim H^1(\fU, G) \\
= & \coprod {H^1(\fU, G)} / \{h_1 \sim  h_2, h_i \in  H^1(\fU_i, G),  i=1,2 \,\,\iff \,\, \exists  \fW: \fU_i < \fW, , \mu_1^*(h_1)=\mu_2^* (h_2) \}.
\end{align}

Nonabelian modifications to the definition of the cocycles and to the definition of the relevant equivalence relation are needed and are implemented at the first step. Most crucially, only for $n=1$, do $Z^n$ and $H^n$ have simple definitions. The cocycle condition now takes the form 
\be\label{cocycle2}
f_{ij} f_{jk} = f_{ik}, \hskip 1cm \mathrm{for}\,\, U_i \cap U_j \cap U_k \neq \emptyset.
\ee and the equivalence relation on the cocycles becomes: $\{f_{ij} \} \sim \{\tilde{f}_{ij} \}, {(i,j) \in N(\fU)}$ if and only if there exists $\{g_i \}_{i \in I} \in C^0(\fU, G)$ such that
\be\label{equiv}
 \tilde{f}_{ij} = g_i^{-1} f_{ij} g_j. 
\ee
The resulting $H^1(\fU, G)$ no longer possesses the structure of a group. 

In practice, one rarely uses the definition for computation, since one hardly needs to first compute  $H^1(\fU, G)$ for \emph{all} open covers $\{ \fU \}$. In stead, one chooses to work with one sufficiently well-behaved open cover for the sheaf and computes its cohomology, which is guaranteed on general ground to be identical to that of $\Sigma$.   

\vskip 1cm

By the above definition, a projective structure on $\Sigma$ clearly defines a class $\xi\in H^1(\Sigma, PSL(2, \C))$. Choosing one atlas of projective coordinate charts $\fU=\{(U_i, z_i)\}_{i \in I}$, the set of transition functions $\{ f_{ij} \triangleq z_i \circ z_j^{-1}: z_j(U_i \cap U_j) \to z_j(U_i \cap U_j)\}_{(i,j) \in N(\fU)}$ defines a 1-cocycle in $Z^1(\fU,PSL(2, \C) )$. The freedom to make local projective coordinate changes in each chart imposes the equivalence relation (\ref{equiv}). And the process of completing the projective atlas by including all compatible projective coordinate charts implements the final step of taking the direct limit (\ref{directlimit}).  

The class $\xi \in H^1(\fU, PSL(2, \C))$ corresponding to a projective structure also has an important additional feature. By definition, it has a nontrivial global section consisting of local homeomorphisms. This is simply given by the local coordinate functions of the projective coordinate covering $\{ z_i \in \Ga_0(U_i, \sC) | (U_i, z_i) \in \fU \} \in \Ga(\fU, \sC(\xi))$. Here $\sC$ (or $\sC(\xi)$) is the sheaf of germs of complex-valued continuous functions (or continuous sections of $\xi$ respectively) on $\Sigma$.  $\Ga_0(U_i, \sC) \subset \Ga (U_i, \sC)$ is the subset of sections over $U_i$, each member of which is a homeomorphism from $U_i$ onto its image in $\C$. 

Much of this section is about characterizing the subset of coordinate classes.

\vskip 1cm

A class $\xi\in H^1(\Sigma, G)$ also has a second geometric interpretation as a flat-$G$-bundle-isomorphism class, in the following natural sense. A $G$-bundle, as is well known, has a local coordinate description. Now, the constancy of the sheaf dictates the constancy of the $G$-valued transition functions in the overlaps between neighboring charts of the fibre bundle. The procedure of taking the equivalence classes of $Z^1(\fU, G)$ removes dependence on the data related to the specific choice of fibre coordinates in each chart, and taking the direct limit removes dependence on the data related to the specific open cover upon which the local product space representation is based. 

However, the terminology ``flat-$G$-bundle-isomorphism class'' may cause some confusion.  As defined in (\ref{equiv}), a flat-G-bundle-equivalence relation is established by the 1-coboundaries of the \emph{constant} sheaf $G$. This allows the possibility for two flat-$G$-bundle-\emph{non}-isomorphic classes to be \emph{isomorphic} as $G$-bundle classes, i.e., they may be equivalent via, for example, a 1-coboundary of the sheaf $\sG_\fb$ of germs of smooth $G$-valued functions.\footnote{If $\Sigma$ possesses the additional structure of a smooth manifold.} Hence caution needs to be taken about this point when we adopt this second geometric interpretation.

\subsection{Coordinate Classes in $H^1(M, PSL(2, \C))$}\label{cc}

Return our discussion from the case of a topological surface $\Sigma$ back to the case of a Riemann surface $M$. The wealth of knowledge on holomorphic line bundles in this context allows a complete characterization of the coordinate classes \cite{Gunning}. We again restrict to $g \geq 2$ (until section \ref{lowg}) and state the result as follows. 

{\bf THEOREM 1} \hskip 0.3cm  Given a Riemann surface $M$ of $g \geq 2$, $\xi\in H^1(M, PSL(2, \C))$ is a coordinate cohomology class \emph{if and only if} both of the following two conditions hold: 
\begin{itemize}
\item[(i)]  $\xi = \rho^*(T)$, for some $T \in H^1(M, GL(2, \C))$; \be\label{ii}\ee
%\vskip -5cm
\item[(ii)] $i^*T =\Lambda \in H^1(M, GL(2, \C)_\fh).$ 
\end{itemize}
Here $GL(2, \C)_\fh$ is the sheaf of germs of complex analytic mappings from $M$ to $GL(2, \C)$. The meanings of $\rho^*, i^*$ and $\Lambda$ are as follows. 

Let $\rho$ be the group homomorphism 
\be\label{rho} \rho: GL(2, \C)  \to  PSL(2, \C) \cong PGL(2, \C) \triangleq GL(2, \C)/Z\ee 
where $Z$ is the center subgroup of $GL(2, \C)$ consisting of nonzero scalar transformations on $\C^2$. $\rho$ induces in the natural way a unique map between the corresponding constant sheaves, which in turn induces
a unique map on the cohomology sets $\rho^*: H^1(\fU, GL(2, \C)) \to H^1(\fU, PSL(2, \C))$.

Let $i: GL(2, \C) \to GL(2, \C)_\fh$ be the natural inclusion of the constant sheaf $GL(2, \C)$  in  $GL(2, \C)_\fh$. $i^*$ is the induced map on cohomology $i^*: H^1(\fU, GL(2, \C)) \to H^1(\fU, GL(2, \C)_\fh)$.

$\Lambda$ is a class that explicitly depends on the holomorphic structure of $M$. 

\subsubsection{Solutions to $\lambda^2=\kappa$, and $\Lambda$}\label{sqrt}

Let $\lambda \in H^1(\fU, \sO^*)$ be a class of holomorphic line bundles satisfying $\lambda^2=\kappa$, where $\kappa$, again, is the class of the canonical line bundle \footnote{Which we also denote as $\kappa$. We do not always distinguish the notation for a bundle and for the class of the bundle.} represented by the cocycle $\{\kappa_{ij}= f'_{ij}(z_j(p))^{-1}, p\in U_i \cap U_j \neq \emptyset\} \in Z^1(\fU, \sO^*)$. Such a class of holomorphic line bundles always exists but is non-unique unless $g=0$. In general, there are $2^{2 g}$ solutions because the Picard variety $P(M)$ is a complex analytic torus of complex dimension $g$. 

For $g \geq 2$, $\lambda$ can be found in the following way. First, the Riemann-Roch thereom gives $c(\kappa)=2 g -2$. One can take arbitrary $g-1$ points $\{p_i, i=1,...,g-1\}$ of $M$, and considers the product of the squares of the corresponding point bundles $\eta=\prod_{i=1,...,g-1} \zeta_{p_i}^2$. Since $c(\eta)=2 g -2$, $\omega= \kappa \eta^{-1}$ is a holomorphic line bundle of $c(\omega)=0$. Therefore $\omega$ is a flat holomorphic line bundle parameterized by $P(M)$, a compact abelian variety of real dimension $2 g$. If we set $\lambda = (\prod_{i=1,...,g-1} \zeta_i) \lambda' $, $c(\lambda')=0$ and $\lambda'^2 = \omega$. Then by the abelian group structure of $P(M)$, we have $2^{2 g}$ solutions to $\lambda'$. We can take any one of these as our choice and form $\lambda= (\prod_{i=1,...,g-1} \zeta_i) \lambda'  $. 

Once we have found a solution class $\lambda$, we further choose a convenient representative cocycle  $\{ \lambda_{0,ij} \}$. Since $\kappa_{ij}(p) = g_i(p) \lambda_{ij}(p) ^2  g_j(p)^{-1} $ for $g_i \in \Ga(U_i, \sO^*)$ we can simply choose our representative to be $\lambda_{0,ij} = g_i(p)^{1/2} \lambda_{ij}(p)) g_j(p)^{-1/2} $. The square-roots here and everywhere else that follows are taken by making an arbitrary choice for each $i$ and then sticking to that particular choice. The resulting $\lambda_{0,ij}$ then obviously satisfies the cocycle condition of $\sO^*$, as well as the equation 
\be\label{root}
\lambda_{0,ij}(p)^2 = \kappa_{ij}(p) = (\del z_i /\del z_j (p))^{-1} 
\ee
 not only as an equation of two classes but as one of two cocycles as well. Note that had we proceeded by directly taking the square-root of $\kappa_{ij}(p)$ in each $U_i \cap U_j \neq \emptyset$, we would not be guaranteed to produce an object that satisfies the cocycle condition as we did here.  

Let
\be\label{Lambda0}
\Lambda_{0, ij}=\left[ \begin{array} {cc}  \lambda_{0, ij}(z_j)  &  \frac{\del}{\del z_j}  \lambda_{0,ij}(z_j) \\ &  \\ 0 &  \lambda_{0, ij}(0,z_j)^{-1}   \end{array} \right].
\ee
Note that for the term involving the derivative of $\lambda_{0,ij}$ in the overlap of the charts, one must be very specific about the coordinate variable with respect to which the derivative is taken. With attention paid to this point, it takes only a very small amount of analysis to see that $\{\Lambda_{0, ij} \}$ satisfies the cocycle condition of the sheaves $SL(2, \C)_\fh) \subset GL(2, \C)_\fh$, given that $\{\lambda_{0, ij}\}$ satisfies the $\sO^*$ cocycle condition.

Now $\Lambda \in H^1(\fU, GL(2, \C)_\fh)$ of (\ref{ii}) is defined to be the class of this cocycle $\Lambda_{0. ij} \in Z^1(\fU, GL(2, \C)_\fh)$.

\subsubsection{Rank-2 Complex Vector Bundles}

Theorem (\ref{ii}) asserts the existence of two related rank-2 complex vector bundles on $M$. The first is the flat $\C^2$-bundle $T \in H^1(\fU, GL(2, \C))$, and the second is the holomorphic $\C^2$-bundle $\Lambda\in H^1(\fU, GL(2, \C)_\fh)$ that essentially is determined\footnote{$\lambda$ and $\Lambda$ are both determined  by $\kappa$ up to $2 g$ $\Z_2$-valued holonomies. } by the canonical line bundle $\kappa$. 

Part (i) asserts that if $\xi\in H^1(\fU, PSL(2, \C))$ comes from a projective structure, then it must be possible to lift it to a class $T \in H^1(\fU, PSL(2, \C))$.  Examples indeed exist of $ H^1(\fU, PSL(2, \C))$ classes that are unable to lift into $H^1(\fU, GL(2, \C))$. The obstruction to satisfying the $GL(2, \C)$-cocycle condition resides in the center $Z$ of $GL(2, \C)$. Such examples provide instances where the $H^1(\fU, PSL(2, \C))$ classes are not coordinate.

Even if $\xi$ can be lifted, it still is not guaranteed to be a coordinate class. Its lift in $H^1(\fU, GL(2, \C))$ must still be related to the canonical line bundle $\kappa$ in the way prescribed by (ii) of (\ref{ii}).  The equivalence relation (between two classes of $H^1(\fU, GL(2, \C)_\fh)$) in (ii), when written out explicitly, reads as follows.  There must exist $\{G_i \in \Ga(\fU, GL(2, \C)_\fh)\}_{i \in I} \in C^0(\fU, GL(2, \C)_\fh) $ such that 
\be\label{ii2}
G_i(p) \cdot \Lambda_{0, ij} (p) =  T_{ij} \cdot G_j(p), \hskip 1cm p \in U_i \cap U_j.
\ee
To put it in plain English, each $g_i(z)$ is a nonsingular $2 \times 2$-complex-matrix-valued function that varies holomorphically with $z_i$ in the open set $U_i$, so that (\ref{ii2}) holds in each $U_i \cap U_j \neq \emptyset$. 
Geometrically, it says that $\xi$ is a coordinate class \emph{if and only if} the flat $GL(2, \C)$ bundle $T$ that is a lift of $\xi$ is isomorphic (in the holomorphic sense) to a holomorphic $GL(2, \C)$ bundle $\Lambda$ that is determined by the canonical line bundle $\kappa$. 
 
As one may suspect and we are about to see, the flat, holomorphic $GL(2, \C)$ bundle $T$ identified by the theorem has a global holomorphic section that consist of local homeomorphisms from $M$ to $\C\sP^1$. It produces a set of local, projective coordinates on $M$ that represents the given projective structure on the Riemann surface.

\subsubsection{The Proof}

Now we outline a complete proof of the theorem. We attempt to make the presentation as concise as possible without sacrificing clarity and soundness.

\emph{only if}   ``$\Longrightarrow$''

Given a coordinate class $\xi$, or equivalently, a  projective structure, we choose a sufficiently representative holomorphic coordinate covering $\fU = \{ U_i, z_i \}_{i \in I}$,  so that $\xi$ is represented by the cocycle of $Z^1(\fU, PSL(2, \C))$ consisting of the transition functions $\{\hz_i \circ \hz_j^{-1}\}$. $\{U_i, \hz_i \}$ is a set of local projective coordinates representing the projective structure, and is compatible with the holomorphic structure of $M$ represented by $\{U_i, z_i\}$. We therefore have $\hz_i(p) = \hu_i \circ z_i (p), p\in U_i$ for some holomorphic function $\hu_i$ defined in the region  $z_i(U_i) \subset \C$ where $\hu' \neq 0$. The proof given in section \ref{sigma} of the existence of the projective structure (in particularly the equations (\ref{c}), (\ref{u}),  (\ref{c2}), and (\ref{connection}), now read in that order) establishes that $\{ (S_2 \hu_i) (z_i (p) ), p \in U_i) \}$ is a projective connection, which we  call $\{ \hh_i \}$.   

Consider once again the set of 2nd order differential equations (\ref{v}), which we reproduce here, with $h_i(z_i)$ replaced by $\hh_i(z_i)$ from the coordinate class $\xi$
\be\label{hv}
2 v''_i(z_i) +  \hh_i(z_i)\cdot v_i (z_i) =0.
\ee
It has two linearly independent holomorphic solutions in each $U_i$, which we group into a 2-vector and call
\be\label{vect}
\hV_i(z_i) = \left[ \begin{array} {c}  \hv_{1i}(z_i) \\   \\ \hv_{2i}(z_i)  \end{array} \right]. 
\ee
Direct computation shows that 
\be\label{tilvect}
\tv_{aj}(z_j (p)) =  \hv_{ai}(z_i (p) ) \cdot  \lambda_{0, ij}(z_i (p)) \hskip 1cm a=1,2, \hskip 0.6cm p\in U_i \cap U_j
\ee
satisfies (\ref{hv}) in the region $U_j$, with the coefficient function  $\hh_j(z_j)$ and, with the derivatives taken with respect to $z_j$ of course. 
 
This is true as a result of a number of key facts. As we evaluate the $\del^2/\del z_j^2$ derivative of $\tv_{aj}(z_j )$, we generate three types of terms  $\hv_{ai}''(z_i  )$,
$\hv_{ai}'(z_i)  \lambda_{0, ij}'(z_i)$, and $\hv_{ai}(z_i) \cdot \lambda_{0, ij}''(z_i) $, which are grouped according to their respective derivative structure\footnote{In writing these terms, we 
have suppressed all factors of powers of $\lambda_{0, ij}(z_i)$}. The second type of terms cancel out, as a result of (\ref{root}). The first type of term returns $\hh_i(z_i)\cdot  \hv_{ai}(z_i)$ after using (\ref{hv}). And the third type of terms generate the product  $\sigma_{2 ji} (z_i) \cdot  \hv_{ai}(z_i)$ involving the canonically associated coboundary $\sigma_{2 ji}$ of (\ref{canonical}), again by (\ref{root}) and by (\ref{schwarzian2}). That these remaining two terms ``conspire'' with $\hh_j(z_j)\cdot \tv_j (z_j)$to satisfy (\ref{hv}) in $U_j$ is just the statement that $\{\hh_i\}$ is a projective connection and hence satisfies (\ref{connection0}) (and more explicitly (\ref{connection})).

$\{ \tv_{aj}(z_j (p)), a=1,2\}$ hence provides in $U_i \cap U_j$ a second set of two independent holomorphic solutions to
$$2 v''_j(z_j) +  \hh_j(z_j)\cdot v_j (z_j) =0, $$ and therefore must be related to $\hV_j(z_j)$ by a constant nonsingular matrix $T_{ji} \in GL(2, \C)$
\be\label{T}
T_{ji} \hV_i(z_i(p)) \cdot  \lambda_{0, ij}(z_i (p)) =  \hV_j(z_j(p)) \hskip 1cm p \in U_i \cap U_j.
\ee
The full transition matrix function between the solution vectors $\hV(z)$ over $U_i$ and $U_j$ is therefore $T_{ji}  \cdot  \lambda_{0, ij}(p) $. It must obviously satisfy the cocycle condition of $GL(2, \C)_\fh$. 
In fact, geometrically (\ref{T}) means that $\{\hV_i\}$ is a global holomorphic section of the rank-2 holomorphic vector bundle defined by the cocycle $\{ T_{ji} \cdot \lambda_{0, ij} \} \in Z^1(\fU, GL(2, \C)_\fh)$. On the other hand, $\lambda_{0, ij}(p)$ is a cocycle of $\sO^*$. Therefore $T_{ji}$ also satisfies the cocycle condition, now of the constant sheaf  $GL(2, \C)$.

We next show that $T_{ji}$ represents a lift of $\xi$ to $H^1(\fU, GL(2, \C))$, i.e.,   $ \rho^*(T) = \xi$. The Wronskian of two independent solutions of (\ref{hv})
\be\label{wron}
W(\hV_i)(z_i) \triangleq \det \left[ \begin{array} {cc} \hV_i(z_i) &  \hV_i'(z_i)  \end{array} \right] 
\ee
is a nonzero constant. As usual, its constancy is shown by taking the derivative with respect to $z_i$ and then use (\ref{hv}); its nonvanishing then follows from linear independency of the solutions.
This implies, in particular, that two independent solutions can not simultaneously take the zero value at the same point. Therefore, treating the 2-vector as the homogeneous coordinates of $\C\sP^1$,  $\hV_i(z_i)$ defines a holomorphic homeomorphism from $U_i$ onto its image in $\C\sP^1$. Holomorphicity is clear. To see it is also homeomorphic, we simply compute
\be\label{ratio}
\frac{\del}{\del z_i} \left(\frac{\hv_{1i}(z_i)}{\hv_{2i}(z_i)} \right)=\frac{W(\hV_i)(z_i)}{\hv_{2i}(z_i)^2}, \hskip 1cm \frac{\del}{\del z_i} \left(\frac{\hv_{2i}(z_i)}{\hv_{1i}(z_i)}\right)=- \frac{W(\hV_i)(z_i)}{\hv_{1i}(z_i)^2}
\ee  
By the above observations, we know at each point, at least one of the two functions $\hw_i(z_i) \triangleq \frac{\hv_{1i}(z_i)}{\hv_{2i}(z_i)}$ and $\hw_i(z_i)^{-1}$ is well-defined.  (\ref{ratio}) then shows that its derivative is nonzero. 

This shows that $\{\hV_i(z_i)\}$ of (\ref{vect}) interpreted as homogeneous coordinates on $\C\sP^1$, or equivalently $\{ w_i(z_i)\} $, represents the projective structure of the coordinate class $\xi$. One can check this by computing $(S_2 w_i)(z_i)$, which, after some algebra, is equal to $\hh_i(z_i)$\footnote{One may still worry about the presence of points in $U_i$ where $w_i$ has simple poles (higher poles are not possible because nontrivial solutions to (\ref{hv}) only have simple zeros). While this is a legitimate concern, the issue does not cause any essential difficulty. All such points must be isolated, and hence can be dealt with simply by refining the cover used around these points and making appropriate projective transformations to move $w_i$ near those points away from $\infty$.}. This is exactly what is meant by the correspondence between a projective structure and a projective connection, as was found out in section \ref{sigma}.   Since the transition formula (\ref{T}) shows that in each $U_i \cap U_j \neq \emptyset$, the transition function $w_j \circ w_i^{-1}$ is given precisely by $\rho(T_{ji})$, we have established that $\rho^*(T) = \xi \in H^1(M, PSL(2, \C))$.
 
To see (ii) of (\ref{ii}), we rewrite (\ref{T}) as
\be\label{T2}
T_{ji} \hV_i(z_i(p))  =  \hV_j(z_j(p)) \cdot  \lambda_{0, ji} (z_i (p)) \hskip 1cm p \in U_i \cap U_j.
\ee
Take the $\del/\del z_i$ derivative on both sides, and after a little bit of algebra, we get 
\be\label{T3}
T_{ji} \del_i \hV_i(z_i) =  \left[ \begin{array} {cc} \hV_j(z_j) & \del_j \hV_j(z_j)  \end{array} \right] \cdot  \left[ \begin{array} {c} \del_i \lambda_{0, ji}(z_i) \\  \\ \lambda_{0, ji} (z_i)^{-1}  \end{array} \right]
\ee
where $\del_k:= \del/\del z_k$. Combining (\ref{T2}) (\ref{T3}) we arrive at 
\be\label{T4}
T_{ji}  \left[ \begin{array} {cc} \hV_i(z_i) & \del_i \hV_i(z_i)  \end{array} \right] = \left[ \begin{array} {cc} \hV_j(z_j) & \del_j \hV_j(z_j)  \end{array} \right]  \cdot \Lambda_{0, ji}.
\ee
This is exactly of the form of (\ref{ii2}), if we substitute
\be\label{g}
g_i (z_i) = \left[ \begin{array} {cc} \hV_i(z_i) & \del_i \hV_i(z_i)  \end{array} \right].
\ee

\vskip 0.5cm
\emph{if} ``$\Longleftarrow$''

The above proves that (i) and  (ii)  of (\ref{ii}) are necessary conditions for $\xi$ to be a coordinate class. They are also sufficient, which we now establish. 

Suppose $$G_i (z_i) = \left[ \begin{array} {cc} g_{11, i}(z_i) & g_{12, i}(z_i) \\  \\g_{21, i}(z_i) & g_{22, i}(z_i)  \end{array} \right], \hskip 1cm i \in I  $$ is a set of nonsingular holomorphic matrix-valued functions such that the equivalence relation (\ref{ii2}) is satisifed, as was assumed by (ii) of (\ref{ii}).  Then as the calculation above (from (\ref{T2}) to (\ref{T4})) showed, 
$$\tG_i (z_i)= \left[ \begin{array} {cc} g_{11, i}(z_i) &  \del_i g_{11, i}(z_i) \\   \\ g_{21, i}(z_i) &\del_i g_{21, i}(z_i)  \end{array} \right], \hskip 1cm i \in I  $$ also satisfies (\ref{ii2}).
Taking the determinant of (\ref{ii2}) we have 
\be\label{det}
    \det \tG_i (z_i) = \det T_{ij} \det \tG_j(z_j).
\ee
Namely $\{ \det \tG_i \}_{i \in I}$ forms a holomorphic global section of the holomorphic line bundle $\{\det T_{ij}\}_{(i,j) \in N(\fU)}$. Since the $T_{ij}$'s are constant matrices, the line bundle has Chern class 0, and by (\ref{chern}) $\det \tG_i (z_i)$ can not have isolated zeros. Either $\det \tG_i (z_i(p))$ vanishes identically on $M$, or it is a nowhere vanishing global holomorphic section of $\det T$ which, therefore, is the trivial line bundle $1 \in H^1(M, \sO^*)$. 

In the latter case, 
$$\left[ \begin{array} {c} g_{11, i}(z_i)  \\  \\g_{21, i}(z_i)   \end{array} \right]$$
provides a good local projective coordinate systme of $U_i$ compatible with the holomorphic structure of $M$, just as we have shown before in the two paragraphs above (\ref{T2}). Furthermore, since $\tG$ satisfies (\ref{ii2}), the associated coordinate class is indeed $\xi = \rho^* (T)$.

To rule out the first case, we assume $\det \tG_i (z_i(p))$ is identically zero for all $i \in I$. Hence we can write
\be\label{nonesense}
\tG_i (z_i) =   \left[ \begin{array} {c} a_i   \\  \\ b_i   \end{array} \right] \cdot  \left[ \begin{array} {cc} g_i(z_i)  &  \del_i g_i(z_i)  \end{array} \right]
\ee
for $a_i, b_i \in \C$. $g_i(z_i)$ is holomorphic and nowhere vanishing in $U_i$ because $G_i(z_i)$ is holomorphic and nowhere singular. Plug this form into (\ref{ii2}), the equality of the first column of the matrices on both sides of that equation then implies
\be\label{nonesense2}
g_i(z_i) \lambda_{0, ij} (z_i) = c_{ij} g_j(z_j)
\ee
for some $c_{ij} \in \C$. This means that the holomorphic line bundles defined by the cocycles $\{c_{ij}\}$ and $\lambda_{0, ij}$ are isomorphic. This is impossible as $\lambda_{0, ij}$ defines a line bundle of Chern class $g-1$, while $\{c_{ij}\}$ defines a line bundle of Chern class 0.

This completes the second part of the proof and the conditions (i) and (ii) of (\ref{ii}) are indeed sufficient.

\vskip 0.5cm

\emph{Final Remarks}  \hskip 0.2cm

An important bonus we derive from this exercise is a quite direct link from the projective connection to the coordinate class. This is not manifest in the statement of the theorem (\ref{ii}), but is otherwise revealed by (\ref{hv}) and (\ref{T}).

A second bonus is the technical observation that things still work if $GL(2, \C)$ is replaced by $SL(2, \C)$ everywhere in Theorem 1 and in the constructions that follow.  This is because
the class $\Lambda$ by its construction (\ref{Lambda0}) is already in $H^1(M, SL(2, \C)_\fh)$. Thus if $T \in  H^1(M, GL(2, \C))$ satisfies the claims of Theorem 1, then so does $T/\det T \in H^1(M, SL(2, \C))$.

\subsection{(Non-)Uniqueness}\label{uniq}

A natural question that has been in one's mind is whether the map from the set of projective structures to the set of coordinate classes is one to one. One may suspect that the answer is negative, as the classes of 
$ H^1(\Sigma, PSL(2, \C)) $  of the constant sheaf provides too crude a ``topological'' classification of the projective structures. In fact,  according to \cite{Gunning},  it is a direct consequence of the Simultaneous Uniformization Theorem \cite{Bers} \cite{Hubbard} that Riemann surfaces of the same underlying topology but of distinct holomorphic structures admit projective structures with the same coordinate class in $H^1(\Sigma, PSL(2, \C))$.

If we once again restrict the discussion to a fixed Riemann surface $M$, this map is injective. In fact, a even stronger statement is true, which we now state as: 

{\bf THEOREM 2} \hskip 0.3cm If two projective structures on the same Riemann surface share the same coordinate class, then not only are they isomorphic, they are also compatible and hence can be coalesced into a single projective structure.

The proof, which is not difficult, is beautiful and somewhat subtle. The version we provide below deviates somewhat from the set of proofs given in \cite{Gunning}, but the underlying idea is still very much the same.

Suppose $\fU_a = \{(U_i, z^{(0)}_{a,i})\}_{i \in I}$ with $a=1,2$ are two sets of projective coordinate charts on the Riemann surface $M$ that produce the same coordinate class $\xi \in H^1(M, PSL(2, \C))$. By Theorem 1 (\ref{ii}) and the second remark after the end of its proof, there exists a lift $T \in  H^1(M, SL(2, \C)) $ of the class $\xi=\rho^* (T)$. We choose a cocycle  $\{T_{ij} \in  SL(2, \C) \}_{(i.j) \in N(\fU)} \in Z^1(\fU, SL(2, \C)) $ representing the class $T$. $\{\rho (T_{ij}) \in PSL(2, \C)\}_{(i.j) \in N(\fU)} \in Z^1(\fU, PSL(2, \C)) $ then represents the coordinate class $\xi$. This means that we can adjust the two sets of projective coordinate maps so that $\fU_a = \{(U_i, z_{a,i})\}_{i \in I}$ with $a=1,2$ have an identical form of transition function on each overlap $U_i \cap U_j \neq \emptyset$. More explicitly, if on $U_i \cap U_j \neq \emptyset$ we have 
\be\label{lift1}
T_{ij}= \left[ \begin{array} {cc} a_{ij} & b_{ij} \\  \\ c_{ij} & d_{ij}  \end{array} \right] \in SL(2, \C), 
\ee then 
\be\label{lift2}
z_{a, i}(p) = \frac{a_{ij} z_{a, j}(p)  + b_{ij} }{c_{ij} z_{a, j}(p) + d_{ij} }.
\ee 
However we must stress  that  for $a=1,2$, even though the fractional linear transformations (\ref{lift2}) have an identical form, the \emph{domains} $z_{a ,j} (U_j) \subset \C $ on which the transformation acts are \emph{a priori different} for $a=1,2$.

We now make the   

\emph{CLAIM} \hskip 0.6cm
For each $a$ of $a=1,2$, there exists a flat complex line bundle $\iota_a \in H^1(\fU, \C^*)$ and a section $f_a = \{(f^1_{a,i} f^2_{a,i})\}_{i \in I} \in \Ga(\fU, \sO(\iota_a T, \C^2))$ of the flat $GL(2, \C)$ bundle $\iota_a T$ such that $z_{a, i}=f^1_{a,i}/ f^2_{a,i}$.\be\label{claim} \, \ee %As this argument is completely parallel for $a=1$ and $a=2$, we suppress the index $a$ temporarily, which should not cause any confusion.  

As we did in section \ref{sqrt}, we choose a set of $g-1$ points $\{p_1, ..., p_{g-1}\}$ on $M$, form the product of the corresponding point bundles $\zeta=\prod_{i=1,...,g-1} \zeta_{p_i}^2$, and represent the canonical line bundle $\kappa$ as $\kappa = \omega \cdot \zeta$ with the help of a flat ($c=0$) holomorphic line bundle $\omega$. In the projective coordinate systems $\{z_{a, i}\}_{i \in I}$ we are currently working with, the canonical line bundle $\kappa$ has its representative cocycles taking a particular simple form
\be\label{kappa}
\kappa_{a, ij} (p) =( \del z_{a, i} /\del z_{a, j} )^{-1}(p) = (c_{ij} z_{a, j}(p) + d_{ij})^2 \hskip 1cm p \in U_i \cap U_j,
\ee
and hence so does the line bundle $\zeta = \omega^{-1} \cdot \kappa $ 
\be\label{zeta}
\zeta_{a, ij} (p) = \omega_{ij}^{-1} (c_{ij} z_{a, j}(p) + d_{ij})^2, \hskip 1cm  \omega_{ij} \in \C^*, \hskip 1cm p \in U_i \cap U_j.
\ee
By definition, $\zeta$ has a global holomorphic section $h$ whose divisor is $\vartheta(h) = \sum_{i=1,...,g-1} 2 \cdot p_i $. Since $h$ only has double-zeros, we can take the square-root of $h$ in each $U_i$, producing the holomorphic local sections $\{g_i \in \Ga(U_i, \sO )\}$. (\ref{zeta}) implies that
\be\label{gsection}
g_{a,i}(p) = \iota_{a,ij} (c_{ij} z_{a, j}(p) + d_{ij}) g_{a,j}(p), \hskip 1cm  \iota_{a, ij} \in \C^*, \hskip 1cm p \in U_i \cap U_j.
\ee 
Again we stress that even though the bundles $\kappa, \zeta, \omega$ and the sections $h, g$ are all the same between $a=1$ and $a=2$,   we can not conclude that, for fixed $i$ and $j$, $g_{a, i}$ and $\iota_{a, ij}$ are the same for $a=1,2$, since the coordinate sections $z_{a, i}$ for $a=1,2$ are a priori related by an arbitrary holomorphic transformation.

If we define 
\begin{align}\label{homocoord}\nonumber
  f^1_{a,i}(p)  & = g_{a,i} (p) \cdot  z_{a, i}(p)\\
  f^2_{a,i}(p)  & = g_{a,i} (p),
\end{align}
then by (\ref{lift2})and (\ref{gsection}), we have 
\begin{align}\label{homocoord2}\nonumber
  f^1_{a,i}(p)  & = \iota_{a,ij} (a_{ij}  f^1_{a,j}(p) + b_{ij} f^2_{a,j}(p)  ) \\
  f^2_{a,i}(p)  & = \iota_{a,ij} (c_{ij}  f^1_{a,j}(p) + d_{ij} f^2_{a,j}(p)  ) 
\end{align}
Since the the coordinate sections $z_{a, i}$ are local homeomorphisms from $U_i$ to $z_{a, i}(U_i) \subset \C^1$, the two holomorphic functions $(f^1_{a,i}(p), f^2_{a,i}(p))$ of (\ref{homocoord}) are linearly independent as functions defined in  $U_i$. Therefore (\ref{homocoord2}) implies $\iota_{a, ij} T_{ij}$ must satisfy the cocycle condition, and hence defines a flat rank-2 complex vector bundle with $(f^1_{a,i}(p), f^2_{a,i}(p))$ as a global holomorphic section. Since both $T_{ij}$  and  $\iota_{a, ij} T_{ij}$ satisfy the cocycle condition, so must $\iota_{a, ij}$, which therefore defines a flat line bundle class $\iota \in H^1(M, \C^*)$. And from (\ref{homocoord}) we see that $z_{a, i}(p) = f^1_{a,i}(p)/ f^2_{a,i}(p)$. Hence our claim (\ref{claim}) has been established.

\vskip 0.5cm

Now consider the matrix-valued holomorphic function $F_i(p)$:
\be\label{matrix}
(F^u\,_a)_i (p) = f^u_{a,i}(p), \hskip 1cm u=1,2; \,\,a=1,2;\,\,i \in I;\,\,p \in U_i.
\ee
It transforms as
\be\label{matrix2}
F_i(p) = T_{ij} F_j(p) \left[ \begin{array} {cc} \iota_{1, ij}  & 0 \\ 0 & \iota_{2, ij} \end{array} \right], \hskip 1cm p \in U_i \cap U_j \neq \emptyset.
\ee
Therefore $\{\det F\} _i =\{\det (F^u\,_a)\}_i  $ is a holomorphic global section of the flat line bundle $\det T \cdot \iota_1 \cdot \iota_2$.  This line bundle has Chern class $0$, hence by (\ref{chern}), the section $\{\det F\}_i$ is either identically zero or nowhere vanishing. But by construction, the second row of $F$ is $F^2\,_a = g_a$, which has simple zeros at the $g-1$ points $p_1, ..., p_{g-1}$. Hence $\det F \equiv 0$. This mean
$z_{1, i}(p)=f^1_{1,i}(p)/ f^2_{1,i}(p)=f^1_{2,i}(p)/ f^2_{2,i}(p)=z_{2, i}(p)$ for all $p \in U_i$ and all $i \in I$. I.e. the two sets of projective coordinate maps are exactly identical.

Thus the two projective structures are not only isomorphic. They are in fact compatible and can be coalesced into a single projective structure. One may wish to pause to appreciate what an amazing uniqueness result this is. 

At the practical level, this result establishes a canonical bijection between $H^{1(c)}(M, PSL(2, \C))$ and $H^0(M, \sO(\kappa^2))$ for the Riemann surface $M$.

\section{Projective Structures at Low Genera}\label{lowg}

Here are three reasons why the $g=0$ and $g=1$ cases deserve additional comments. 
\begin{itemize}
\item[(i)]  Unlike $g \geq 2$ in which case $H^1(M, \sO(\kappa^2))=0$ and hence the canonically associated cocycle $\sigma_2$ must be trivial, $H^1(M, \sO(\kappa^2))$ does not vanish for $g=0,1$.  
\item[(ii)] The point bundles played a major role in the constructions of section \ref{cc}. And the ``point bundle'' $\zeta$ at $g=0$ is qualitatively different from the point bundles of $g \geq 1$ surfaces.
\item[(iii)]$g=1$ is the only case that admits affine structures in addition to admitting projective structures.  
\end{itemize}

The simplicity of the $g=0$ and $g=1$ cases (manifested, for example, by the existence of nontrivial conformal Killing vector fields in these two cases)
allows them to be analyzed explicitly by direct, elementary methods.

\subsection{$g=0$}\label{CP}

Since $c(\kappa) =-2$, the set of infinitesimal deformations to the complex structure is vacuous. In fact, the Uniformization Theorem \cite{FK} \cite{Hubbard} implies that any two complex manifolds of the topology of the sphere are biholomorphic to each other. This does \emph{not} mean that, on a topological sphere, there exists only one holomorphic structure. On the contrary, there exists a continuum infinity of distinct holomorphic structures. 
\footnote{They are distinct in the sense of being \emph{mutually incompatible}. If $\fU_{(a)} =\{(U_i, z_{i, (a)})\}_{i \in I}$, $a=1,2$ are two such holomorphic coordinate coverings supported by the same underlying open cover $\{U_i\}_{i \in I}$, mutual incompatibility means $f_{(12),i} (z) \triangleq z_{i, (1)} \circ z_{i, (2)}^{-1} (z), z \in z_{i,(2)}(U_i) $ is not holomorphic.}  But it does mean that they are all equivalent in the sense of being biholomorphic to one another.\footnote{Namely, for any two such structures, there exists a map from the space to itself such that when expressed in terms of the local holomorphic coordinates associated to the two structures, the map is a holomorphic homeomorphism in the appropriate domain.} An observer on one of these holomorphic spheres can not tell the difference of her world with any of the other holomorphic worlds if she is only allowed to carry out analysis that are exclusively meromorphic. She will however be able to discern the difference if more probing, non-meromorphic mathematical instruments are introduced.

Within this ``meromorphically unique'' holomorphic world, it is self-evident that a projective structure exists. This is the structure that gives the letter ``P'' to the terminology $\C\sP^1$. Even though by the Serre duality $H^1(\C\sP^1, \sO(\kappa^2)) \cong \Ga(\C\sP^1,\sO(\kappa^{-1}) ) \cong \C^3$, $\sigma_2$ clearly vanishes when computed in the standard $\{U_+, z_+\} \cup \{U_-, z_-\}$ coordinates which are identified by $z_+ z_-=1$ near the equator. 

As was already stated in section \ref{uniformization}, this projective structure is unique (within the above context of biholomorphic equivalence). This was determined in section \ref{uniformization} by exploiting the one-one correspondence between the projective structures and the projective connections, and $\Ga(\C\sP^1,\sO(\kappa^2))=0$. However, this is also a direct consequence of the  well-known fact that the group $\Aut (\C\sP^1)$ of all holomorphic automorphisms of $\C\sP^1$ is precisely the group of all M\"{o}bius transformations (i.e. projective linear transformations). This means that any two projective structures on $\C\sP^1$ subordinate to the same holomorphic structure must be projectively related, and therefore can be combined within a single projective structure.

\subsection{$g=1$}\label{torus}

The standard description of a $\sfT^2$ as $\C/\{\Z \times \Z_\tau\}$ by the parametrization $z \sim z+1 \sim z+\tau$ represents an affine structure on the  $\sfT^2$ in addition to corresponding to a projective structure. 
It is clear that a compact Riemann surface admits an affine structure only if its topology is a torus. In the affine coordinates, the transition functions between the patches take the affine form 
$$z_i(p) = a_{ij} z_j(p) + b_{ij}, \hskip 1cm a_{ij} \neq 0$$
the canonical line bundle $\kappa$ is represented by a cocycle 
$$\kappa_{ij} = (\del z_i/\del z_j)^{-1} = a_{ij}^{-1}$$ which is in the constant subsheaf $\C^* \subset \sO^*$. Therefore $\kappa$ must be a flat holomorphic line bundle with $c(\kappa)=0$. On the other hand, we know from the Riemann-Roch theorem  and the Serre duality that 
$$c(\kappa) = 2 g -2.$$
Hence a surface must have $g=1$ to admit an affine structure.  

As already explained in the footnote \footref{torusline}, at $g=1$ we have  $\kappa =1 \in H^1(\sfT^2_\tau, \sO^*)$. Therefore the spaces of incompatible affine structures and of incompatible projective connections are, respectively, 
\be\label{prediction}
\Ga(\sfT^2_\tau, \sO(\kappa)) = \Ga(\sfT^2_\tau, \sO)= \C, \hskip 1cm  \Ga(\sfT^2_\tau, \sO(\kappa^2)) = \Ga(\sfT^2_\tau, \sO)= \C. 
\ee
We thus have a total of one-complex-parameter-worth of affine and projective structures including the canonical structures represented by the standard coordinate $z$ above. We now explicitly identify them as follows.

Given the explicit algorithm we outlined in section \ref{PS} that implements the one-one correspondence between the affine/projective structures and the affine/projective connections, we only need to solve (\ref{u}) in the projective case, and its counterpart with the $S_1$ operator replacing the $S_2$ in the affine case. In principle, these equations should be solved patch by patch on $\sfT^2_\tau$, and the new structures are encoded in the transition functions between these patch-wise solutions. In practice, it is technically simpler to work with the covering space $\C$. The affine and projective connections, both being global sections of $\sO$ over $\sfT^2_\tau$, are simply constant complex numbers. So they can be trivially extended to $\C$. The resulting solutions $u(z)$, however, are not expected to be invariant under the covering transformations. Their changes under the covering transformations encode the transition functions between the original coordinate patches on $\sfT^2_\tau$, which in turn represent the new structures we are solving for.   

\vskip 0.8cm

\subsubsection{The Affine Structure Moduli Space $\cA_{g=1} = \Lambda _{\cM_{g=1}}/ \Z_2$}
We first look for the affine structures. As outlined above, we solve
\be\label{affine}
(S_1 u) (z) = h 
\ee
on the complex plane with a constant $h$. This equation is simply 
$$u''(z)- h \, u(z) =0$$ 
with its general solution be 
\be\label{affinesol}
u(z)=a\, \exp(h\, z) +b  \hskip 0.5cm \mathrm{if} \,\,h \neq 0; \hskip 1.3cm  u(z)=a\, z + b \hskip 0.5cm \mathrm{if} \,\, h=0
\ee
with $a, b \in \C$ arbitrary integration constants. 

Since we are solving for the \emph{new} affine \emph{structures}, only the non-affine piece of the solution is relevant. Hence we can take simply
\be\label{affinesol0}
u(z, h)=\exp(h\, z), \hskip 1.5cm  h \neq 0.
\ee
Different choices of the integration constants only produce equivalent structures.

Now as a consistency check, under the covering transformations 
\be\label{period}
z \to z+1 \hskip 1cm \mathrm{and}  \hskip 1cm z \to z+\tau,
\ee
 the solution $u$ transforms as
\be\label{change}
u \to e^h \,u \hskip 1.15cm \mathrm{and}  \hskip 1cm u \to e^{\tau h} \,u ,
\ee
which indeed are affine transformations. Hence the solutions (\ref{affinesol0}) do define affine structures. 

It is clear that the solutions (\ref{affinesol0}) of different values of $h$ are not linearly related, hence any two of these affine structures are not mutually compatible. On the other hand, it is also clear that the pair of structures defined by $h$ and $-h$ does not have any intrinsic difference. Indeed the map $z \to -z$ is an isomorphism between these two incompatible affine structures. Therefore we should impose the equivalence relation
\be\label{equivalenceall}
h \sim_{\,_{g=1}} - h.
\ee
This is a concrete example manifesting the difference between compatible and isomorphic structures, as we discussed in subsection \ref{orbifold} \footnote{Albeit in the context of projective structures. The underlying idea clearly also applies to the case of affine structure. See footnote \ref{affinevsproj}}. The reason that here we have an equivalence relation over all of $\cM_{g=1}$ even away from the orbifold loci $\Delta_1= \{A, B\}$ (which we will analyze below) is due to a special feature at $g=1$, namely that all tori has a discrete ``parity'' automorphism represented by $z \rightarrow -z$.  

It is straightforward to determine the transformation of the affine-structure parameter $h$ under the modular transformations. Under 
\be\label{plus1}
\tau \to \tau+1,
\ee 
the affine-structure modulus $h$ must be invariant in order to preserve the ``multiplicative lattice''  of (\ref{change}) of the non-canonical affine structure before and after (\ref{plus1}). Under 
\be\label{inver}
\tau \to -1/\tau, 
\ee 
Preservation of the ``multiplicative lattice'' of (\ref{change}) requires that
\be\label{htransf}
h \to \pm \tau \cdot h.
\ee 
That a factor of $\tau$ must arise can also be seen by noting that, under the modular transformation (\ref{inver}), the new uniformizing holomorphic coordinate is, up to a constant shift (and a sign which we will come to next), $z'=z/\tau$. Hence to preserve the affine structure of (\ref{affinesol0}), $h$ must acquire a factor of $\tau$. 

The $\pm$ ambiguity is to be expected, from (\ref{equivalenceall}), and it can not be resolved by continuity.  
Since there is no intrinsic difference between the two affine structures defined by $h$ and $-h$, when we return to the same holomorphic structure along a path in the Teichm\"{u}ller space but with a modular transformation, we can not decide which sign to take based on the requirement that the affine structures should match. One might hope that it may be possible to impose ``by hand'' a notion of continuity by for example choosing the two points related by the modular transformation arbitrarily close in the Teichm\"{u}ller space (this can be done by choosing the two points in question to sit very close to a fixed point). But a large diffeomorphism on the torus always needs to be made in order to match its complex coordinates corresponding to these two points regardless how close or far they are separated. This  carries its own set of discrete choice (indeed, a sign choice, once again) and invalidates the possibility of a selection rule by continuity.  

$\cM_{g=1}$ has two orbifold points $A$ and $B$ of order 2 and order 3 respectively, which can be chosen to be situated at $\tau_A=i$ and $\tau_B=e^{i \pi/3}$ in terms of the standard $\tau$-coordinate on the Teichm\"{u}ller space $\cT_{g=1}$. They correspond to particular symmetric tori that admit additional discrete automorphisms. Even though by changing to new local coordinates $\tilde{\tau}_1= (\tau-i)^2$ and $\tilde{\tau}_2= (\tau-e^{i \pi/3})^3$ around $A$ and $B$ one sees that $\cM_{g=1}$ is in fact isomorphic to the complex plane $\C$, the orbifold singularities in the original $\tau$-coordinate do generate additional subtleties. 

As we explained in subsection \ref{orbifold} and stated more explicitly in item (IV) of subsection \ref{summary} \footnote{Again, as in footnote \ref{affinevsproj}, albeit originally stated in the context of projective structures, the underlying idea clearly also applies to the case of affine structure.}, we expect that on such symmetric surfaces, certain incompatible affine structures are nonetheless isomorphic. This now can be seen explicitly from (\ref{affinesol0}). At the order 2 orbifold singularity $A$ represented for instance by $\tau_A=i$, we have the equivalence relation
\be\label{equivalenceA}
h \sim_{\,_A} i \cdot h.
\ee
This is duo to the additional automorphism $$z \rightarrow i \cdot z$$ that exists for a torus with $\tau=i$.
At the order 3 orbifold singularity $B$ represented for instance by $\tau_B=e^{i \pi/3}$, we have the equivalence relation
\be\label{equivalenceB}
h \sim_{\,_B} e^{i \pi/3} \cdot h,
\ee
which is a result of the additional automorphism $$z \rightarrow  e^{i \pi/3} \cdot z$$ that exists  for a torus with $\tau=e^{i \pi/3}$.

Note that the identifications (\ref{equivalenceA}) and (\ref{equivalenceB}) are performed to $\Ga(\sfT^2_{\tau=i}, \sO) =\C$ and $\Ga(\sfT^2_{\tau=e^{i \pi/3}}, \sO) =\C$, in addition to the overall identification (\ref{equivalenceall}) which must be done everywhere in $\cM_{g=1}$. If we first neglect the global identification (\ref{equivalenceall}) and focus only on (\ref{equivalenceA}) and (\ref{equivalenceB}), then these identifications seem particularly natural given that (a) the proper local, complex analytic coordinates near the orbifold points $A$ and $B$ are $\tilde{\tau}_1= (\tau-i)^2$ and $\tilde{\tau}_2= (\tau-e^{i \pi/3})^3$; and (b) the second equation of (\ref{change}) suggests that the proper affine structure modulus should make corresonding adjustment when adopting the new, lcoal complex structure modulus $\tilde{\tau}_1, \tilde{\tau}_2$. With only the identifications (\ref{equivalenceA}) and (\ref{equivalenceB}) in place, we arrive at a bundle $\Lambda _{\cM_{g=1}}$ that has a copy $\C$ as its fiber over $\cM_{g=1} \backslash \{A, B\}$, a $\C /\Z_4$ fiber over $A$, and a $\C /\Z_6$ fiber over $B$. This bundle appear to be related to the Hodge bundle \cite{HM}, a relationship that we would like to understand better.

To conclude this part, we have found a one-complex-parameter-family of mutually incompatible affine structures on $\sfT^2_\tau$, as was predicted by the first half of (\ref{prediction}). Taking into account the additional discrete automorphisms at the two orbifold points $A$ and $B$, these affine structures are parameterized by the bundle $\Lambda _{\cM_{g=1}}$ over $\cM_{g=1} \cong \C$. The global existence of the discrete ``parity'' automorphism for every $g=1$ Riemann surface then dictates that we must take the $\Z_2$ orbifold (\ref{equivalenceall}) of this bundle, which has its fixed points corresponding to the canonical affine structure on every $g=1$ Riemann surface. The overall affine structure moduli space $\cA_{g=1}$ may therefore be denoted as $\Lambda _{\cM_{g=1}}/\Z_2$, a bundle with $\C/\Z_2$ fibers over $\cM_{g=1} \backslash\{A, B\}$, a $\C/\Z_4$ fiber over $A$, and a $\C/\Z_6$ fiber over $B$.

\vskip 0.8cm

\subsubsection{The Projective Structure Moduli Space $\cP_{g=1} = T^*_{(1,0)} \cM_{g=1}$}

Moving on to the projective structures, we need to solve 
\be\label{proj1}
(S_2 u) (z) = \th 
\ee
on the complex plane with a constant $\th$. For reasons that will become clear momentarily, we set 
\be\label{thh}
\th=-h^2/2
\ee and solve in stead
\be\label{proj2}
(S_2 u) (z) = -h^2/2. 
\ee
The general solution is easily found out to be
\be\label{projsol}
u(z)=\frac{1}{a\, \exp(h\, z) +b}+c  \hskip 0.5cm \mathrm{if} \,\,\,h \neq 0; \hskip 1.3cm  u(z)=\frac{a z+b}{c z+d} \hskip 0.5cm \mathrm{if} \,\,\, h=0
\ee
Again keeping only the non-projective ``nucleus'' of the solutions in order to identify the new projective structures, we have simply  
\be\label{projsol0}
u(z, h)=\exp(h\, z), \hskip 1.5cm  h \neq 0,
\ee
which is exactly identical to (\ref{affinesol0}). This, in large measure, is  to be expected. After all, an affine structure is by itself a projective structure. What this small calculation has additionally confirmed is that,
to each new projective structure $\th \neq 0$, there are precisely two affine structures (the two solution to (\ref{thh})) subordinate to it. These two (equivalent but incompatible) affine structures are now compatible as projective structures because $u(z, h) \cdot (z, -h) =1$. 

Under the modular transformation (\ref{inver}),  the change of the affine-structure modulus $h$ (\ref{htransf}) combined with the subordination map (\ref{thh}) requires that the projective-structure modulus must transform according to
\be\label{thtransf}
\th \to \tau^2  \cdot \th.
\ee
The sign ambiguity cancels out because of the two-to-one relation between the affine and projective structure parameters in (\ref{thh}). 
The deeper reason of course is that, now different values of $\th$ correspond to inequivalent projective structures, and therefore the continuity of the projective structure under modular transforms alone forbids in the projective case, any one-to-many relation like that of (\ref{htransf}).

(\ref{thtransf}) implies that $\th \cdot \d \tau$ is invariant under (\ref{inver}). This is in exact agreement with our expectation from section \ref{uniformization} that the projective-structure modulus acts as the fiber coordinate of $T^*_{(1,0)}\mathcal{M}_{g=1}$ over the moduli space $\mathcal{M}_{g=1}$, at least away from the two orbifold points $A$ and $B$.

The same simple reasoning also immediately implies that $\th \cdot \d \tau$ is preserved by the orbifold actions at the orbifold points $A$ and $B$, which we now see explicitly. The order 2 orbifold point $A$, represented by for instance $\tau_A = i$, is the fixed point of the transformation (\ref{inver}), under which 
\be\label{inver1}
\d \tau \to  \d \tau / \tau_A^2 = - \d \tau
\ee
at $\tau=\tau_A = i$, and the transformation of $\th$ at  $\tau=\tau_A = i$ is, by (\ref{htransf})
\be\label{inver2}
\th \to  \tau_A^2  \cdot \th = - \th.
\ee
The order 3 orbifold fixed point $B$, represented by for instance $\tau_B=e^{i \pi/3}$, is the fixed point of the combined transformation:
\be\label{combine}
\tau \to 1 - 1 / \tau,
\ee
under which, 
\be\label{inver1}
\d \tau \to  \d \tau / \tau_B^2 = e^{-2 i \pi/3} \cdot \d \tau
\ee
at $\tau=\tau_B=e^{i \pi/3}$, and  
\be\label{combine2}
\th \to  \tau_B^2 \cdot \th = e^{2 i \pi/3} \cdot \th.
\ee

\vskip 1cm

\subsubsection{Conclusions at $g=1$}
\begin{itemize}
\item[(I)] On each $\sfT^2_\tau$, we have found the expected one-complex-parameter-family of (incompatible) affine structures $\mathcal{A}^{(0)}_\tau =\C_h$ as well as the expected one-complex-parameter-family of (incompatible) projective structures $\mathcal{P}^{(0)}_\tau=\C_\th$. Their relation to the canonical affine structure $\mathcal{A}_\tau(h=0)$ is given by (\ref{affinesol0}), and the affine-structure modulus $h$ is related to the projective-structure modulus $\th$ by the subordination map (\ref{thh}).  

\item[(II)] At the order-2 orbifold locus $A$ and the order-3 orbifold locus $B$ of the genus-1 complex structure moduli space $\cM_{g=1}$, additional discrete automorphisms arise on the corresponding tori. As a result, affine structures at $A$ related by (\ref{equivalenceA}) and at $B$ related by (\ref{equivalenceB}), become isomorphic at $A$ and at $B$ respectively. Hence the fiber $\mathcal{A}_A$ at $A$ reduces to a copy of $\C/\Z_4$, and the fiber $\mathcal{A}_B$ at $B$ to a copy of $\C/\Z_6$. We denote the resulting bundle, with the fiber remaining $\C$ over the rest of $\cM_{g=1}$ (i.e. over $\cM_{g=1} \backslash \{A, B\}$) by $\Lambda_{\cM_{g=1}}$. This bundle appears to be related to the Hodge bundle over $\cM_{g=1}$, something we would like to understand better. 

\item[(III)] Every $\sfT^2_\tau$ has the ``parity'' automorphism (i.e. $z \to -z$ in the uniformizing coordinate). Hence the two affine structures in $\mathcal{A}^{(0)}_\tau$ related by $h \to -h, h\neq 0$ have no intrinsic difference and are equivalent. This means that the genuine affine structure moduli space $\cA_{g=1}$ is a $\Z_2$ quotient of $\Lambda_{\cM_{g=1}}$. This $\Lambda_{\cM_{g=1}}/ \Z_2$ bundle has its generic fiber a copy of $\C/\Z_2$, a non-generic $\C/\Z_4$ fiber at $A$, and a non-generic $\C/\Z_6$ fiber at $B$.

\item[(IV)] The naive affine structure modulus $h$ is related to the projective structure modulus $\th$ by the two-to-one subordination relation (\ref{thh}). As a result, no overall $\Z_2$ quotient is necessary in the projective structure moduli space, and the generic fiber remains $\cP_\tau = \cP^{(0)}_\tau = \C_\th$. At the orbifold points $A$ and $B$, we have the non-generic fibers $\cP_A=\cP^{(0)}_A/ \Z_2 = \C/\Z_2$ and $\cP_B=\cP^{(0)}_B/ \Z_3 = \C/\Z_3$. 
    
    If we work with the local analytic coordinates $\tilde{\tau}_1= (\tau-i)^2$ and $\tilde{\tau}_2= (\tau-e^{i \pi/3})^3$ near the two orbifold points so that $\cM_{g=1}$ is analytically isomorphic to $\C$, then the projective structure moduli space is exactly identified with the complex analytic $T^*_{(1,0)}\cM_{g=1}$.

\end{itemize}

\section{$\dim_\C H^1(\Sigma, PSL(2, \C)) $}\label{highg}

Explicit computations like those carried out in section \ref{lowg} still seem feasible at moderate values of the genus. But as we go to significantly larger values of $g$, more power technologies seem indispensable for the analysis of the orbifold loci $\Delta_g \subset \cM_g$. One also hopes to study the projective structures on degenerate Riemann surfaces, which correspond to measure-zero boundaries of properly compactified moduli spaces $\bar{\cM}_g$ (or $\bar{\cM}_{g,n}$ if with punctures) \cite{HM} \cite{FriedanShenker}.  This also requires more advanced algebraic geometry technologies. In absence of these, we switch track to make some semi-quantitative comments. 

The canonical bijection established in section \ref{uniq} equips $H^{1 (c)}(M, PSL(2, \C))$ with the structure of a complex analytic manifold of complex dimension $3 g -3$. By computing the dimensionality of the space $H^1(\Sigma, PSL(2, \C))$, we will acquire a semi-quantitative understanding of how special the coordinate classes of a Riemann surface are.  

\vskip 0.6cm

For this computation, we note that
for a manifold $M$,  there is a canonical bijection between sets:
\be\label{bijec} 
\varpi_0: H^1(M, G) \to \Hom (\pi_1(M, x_0), G))/G.
\ee  
Here $G$ is an arbitrary group, not necessarily abelian. $H^1(M, G)$, as before,  is the cohomology set of the constant sheaf $M \times G$ (where, again, we must take the discrete topology of $G$ in forming the product topology). $\Hom (\pi_1(M, x_0), G))$ is the set of homomorphisms from $\pi_1(M, x_0)$ to $G$, $\,\,x_0 \in M$ is an arbitrary base point. If $\chi \in \Hom (\pi_1(M, x_0), G))$, then
\be\label{homo}
\chi: \pi_1(M, x_0) \to G, \,\,\,\,\,\chi(\fl_1 \ast \fl_2) = \chi(\fl_1) \chi(\fl_2)  
\ee
where $\fl_{1,2}$ are path-homotopy classes of loops based at $x_0$. $\Hom (\pi_1(M, x_0), G))$ admits a $G$-action by conjugation. It sends $\chi (\fl)$ to $\chi^g (\fl) = g \chi (\fl) g^{-1}$. $\chi^g$ again satisfies (\ref{homo}), hence is in $\Hom (\pi_1(M, x_0), G)$. $\til{\chi} = \chi^g$ for some $g \in G$ is an equivalence relation in $\Hom (\pi_1(M, x_0), G))$, and $\Hom (\pi_1(M, x_0), G))/G$ is the set of equivalence classes.

There are two ways to view (\ref{bijec}) that hopefully make it seem familiar. On the one hand,  what (\ref{bijec}) says ``physically'' is that the flat $G$-bundles on a space $M$ are \emph{bijectively} classified by the conjugacy classes of the $G$-valued-Wilson lines on $M$. On the other hand, mathematically, it may be viewed as a nonabelian generalization of the Universal Coefficient Theorem for Cohomology in the case of $n=1$
\be\label{UCT}
0 \to \Ext(H_{n-1}(C),G) \to H^n(C, G) \to \Hom(H_n(C), G) \to 0.
\ee
(\ref{UCT}) holds for a chain complex $C$ of free abelian groups, and an arbitrary \emph{abelian} group $G$. If we set $n=1$ in (\ref{UCT}),  and recall that for an abelian group $G$, $\Hom(\pi_1(X), G)/G  \cong \Hom(\pi_1(X), G) \cong \Hom(H_1(X), G)$, and that $\Ext(H, G) = 0$ when $H$ is free, then we immediately recognize (\ref{bijec}) as a nonabelian generalization of  (\ref{UCT}).

\vskip 0.6cm

For us, $G=PSL(2, \C)$ is a complex analytic Lie group of complex dimension 3. We expect both $H^1(M, G)$ and $\Hom (\pi_1(M, x_0), G))/G$  (with $G=PSL(2, \C)$) to have the structure of a complex manifold, and we expect $\varpi_0$  to be promoted to a homeomorphism at least. So we elect to compute the complex dimension of $\Hom (\pi_1(M, x_0), G)/G$ (with $G=PSL(2, \C)$) in stead.

The fundamental group of a genus $g$ Riemann surface is well-known. It is isomorphic to the quotient of the free group on $2 g$ generators by the least normal subgroup containing the element 
$$ [a_1, b_1][a_2, b_2]\cdot \cdot \cdot  [a_g, b_g]$$
where $[a,b] \triangleq a b a^{-1} b^{-1}$. In other words, we have $2 g$ generators, constrained by a single relation. Hence setting in (\ref{bijec}) $G=PSL(2, \C)$, which is a complex 3-dimensional complex analytic Lie group, we see that 
\be\label{dim1}
\dim_\C \{ \Hom [\pi_1(M, x_0), PSL(2, \C)] \}= 3 \cdot (2 g -1) \ee
Hence   
\begin{align}\label{dim2} \nonumber
   & \dim_\C \{ H^1(M, PSL(2, \C))\} \\ \nonumber
 =\,\,\,\,  &  \dim_\C \{ \Hom [\pi_1(M, x_0), PSL(2, \C)]/PSL(2, \C)\} \\
 = \,\,\,\, & 3 \cdot (2 g -1) -3 \,\,\,\, = \,\,\,\, 6 g -6 
\end{align}
This is the same complex dimension as $T^*_{(1,0)} \mathcal{M}_g$.

The bijection relation (Theorem 2) from section \ref{uniq} predicts  $H^{1 (c)}(M, PSL(2, \C)) \cong  \cP_M \cong H^0(M, \sO(\kappa^2))$, which gives $H^{1 (c)}(M, PSL(2, \C))$ the structure of a complex analytic manifold of complex dimension $3 g-3$. Assuming this is the same structure as the one induced from its inclusion in $H^1(\Sigma, PSL(2, \C))$, we conclude that $H^{1 (c)}(M, PSL(2, \C))$ is a middle-dimensional subspace of $ H^1(\Sigma, PSL(2, \C))$.

\vskip 1cm

\vskip 1cm

\noindent{\it Acknowledgments}  \hskip 0.5cm The author wishes to thank his colleagues at UESTC for encouragements. He also wishes to thank the university staff for administerial support.

\appendix
%\section{Uniformization Theorem}\label{uniformthm}

%\section{$\Aut(\C\sP^1)$ and $\Aut(\C)$}

\section{The Serre Duality Theorem}\label{SDT}

We only need a basic version of the Serre duality theorem in this note. We state it as follows.

\emph{Serre Duality} \hskip 0.4cm For any holomorphic line bundle $\xi \in H^1(M, \sO^*)$ on a compact Riemann surface $M$, there is a canonical duality between the complex vector spaces $H^1(M, \sO(\xi))$ and $H^0(M, \sO^{1,0} (\xi^{-1}))$.

In very broad brush strokes, this statement can be established by combining two natural ideas with some very careful analysis. The two ideas are:

(i) The Dolbeault-Serre sequence
\be\label{DolbeaultSerre}
0 \to \sO (\xi) \xrightarrow{\,\,i\,\,} \sE^{0,0} (\xi) \xrightarrow{\,\,\dbar\,\,}  \sE^{0,1} (\xi) \to 0 \ee
provides a fine resolution to the sheaf $\sO(\xi)$. Here $\sE^{p,q} (\xi)$ stands for the sheaf of germs of $(p,q)$-\emph{symmetric tensor} \footnote{Of course, for $(p,q)=(1,0), (0,1)$, and $(1,1)$, these are the same as the sheaf of germs of smooth $(1,0)$-forms, of smooth $(0,1)$-forms, and of smooth $(1,1)$ forms, respectively.}-valued 
\emph{smooth} sections of the holomorphic line bundle $\xi$, and $\dbar: \sE^{0,0} (\xi) \to \sE^{0,1} (\xi)$ defines a sheaf homomorphism because the line bundle $\xi$ is holomorphic. 

Noting that $\sE^{0,0} (\xi)$ and $\sE^{0,1} (\xi)$ are fine sheaves,  performing the standard diagram-chasing to the short exact sequence of cochain group complexes resulting from (\ref{DolbeaultSerre}) immediately yields 
\be\label{fineresol}
H^1(M, \sO(\xi)) \cong \Ga(M,  \sE^{0,1} (\xi)) /\dbar \Ga(M, \sE^{0,0} (\xi))
\ee

(ii) The natural bilinear pairing 
\be\label{pair1}
\Ga(M,\sE^{0,1} (\xi) ) \times \Ga(M,\sE^{1,0} (\xi^{-1}) ) \to \C 
\ee
given by
\be\label{pair2}
\int_M \phi \wedge \psi ,
\ee with $\phi \in \Ga(M,\sE^{0,1} (\xi) )$ and $\psi \in \Ga(M,\sE^{1,0} (\xi^{-1}) )$, vanishes when being restricted to 
\be\label{pair3}
\dbar \Ga(M,\sE^{0,0} (\xi) ) \times \Ga(M,\sO^{1,0} (\xi^{-1})).
\ee It hence induces a bilinear pairing on
\be\label{pair4}
\Ga(M,  \sE^{0,1} (\xi)) /\dbar \Ga(M, \sE^{0,0} (\xi))  \times \Ga(M,\sO^{1,0} (\xi^{-1}))  \to \C,
\ee 
which, by (\ref{fineresol}), is a bilinear pairing on
\be\label{pair5}
 H^1(M, \sO(\xi)) \times H^0(M, \sO^{1,0} (\xi^{-1})) \to \C.
\ee 

To demonstrate that the canonical pairing of (\ref{pair5}) is a duality requires showing that it is nonsingular. This takes some very careful analysis which can roughly be separated into three steps. 

(a) To show that (\ref{fineresol}), which is a quotient between two infinite dimensional spaces,  is actually of \emph{finite} dimension. This is done by introducing proper norms on the spaces of various tensor-valued global sections of $\xi$ to topologize these spaces and showing the resulting quotient of (\ref{fineresol}) is \emph{locally compact}.

(b) To use (\ref{pair2}) as the bridge between  $ H^1(M, \sO(\xi))$ and  $H^0(M, \sO^{1,0} (\xi^{-1}))$, one needs to be able to lift linear functions on the quotient (\ref{fineresol}) to (continuous) linear functionals on $\Ga(M,  \sE^{0,1} (\xi))$. This is done by showing that $\dbar \Ga(M, \sE^{0,0} (\xi))$ is a \emph{closed} subspace of $\Ga(M,  \sE^{0,1} (\xi))$. 
 
(c) The dual space of $\Ga(M,  \sE^{0,1} (\xi))$ is the space of $(1,0)$-form-valued \emph{distributional} sections of $\xi^{-1}$, $\Ga(M, \cD^{1,0}(\xi^{-1})) $. Once the lift of part (b) is proven possible, one needs to show that the subset of distributions that vanish identically on  $\dbar \Ga(M, \sE^{0,0} (\xi))$ is given precisely by the ($(1,0)$-form-valued) \emph{holomorphic} sections of $\xi^{-1}$.

\section{The Riemann-Roch Theorem}\label{RRT}

The version of the Riemann-Roch theorem we need is quite standard. It states that

For any holomorphic line bundle $\xi \in H^1(M, \sO^*)$ on a compact Riemann surface $M$ of genus $g$, 
\be\label{RR}
\dim H^0(M, \sO(\xi)) - \dim H^1(M, \sO(\xi)) - c(\xi) =1-g.
\ee $c(\xi)$ is the integer-valued Chern class of $\xi$. 

Using the Serre duality theorem of Appendix \ref{SDT}, this is equivalent to

\be\label{RR2}
\dim \Ga(M, \sO(\xi)) - \dim \Ga(M, \sO(\kappa \xi^{-1}))  - c(\xi) =1-g. \ee

Its proof can be found in any standard text on Riemann surfaces. We merely point out that the form of the Riemann-Roch theorem from Joe's classic text \cite{Polchinski} (5.3.22) corresponds to taking $\xi =\kappa^n$,  the line bundle of an integer power of the canonical line bundle $\kappa$. These are just the holomorphic line bundle of symmetric tensors of type $(n,0)$, the tensors being ``contravariant'' if $n<0$.  The holomorphic sections of $\xi$ and $\kappa \xi^{-1}$ then correspond to the classical solutions of the general chiral $(b_{(n)}, c^{(n-1)})$ ghost field system, i.e. the ghost zero-modes. Off-shell, the $b_{(n)}$ and $c^{(n-1)}$ ghost fields take their respective values in sufficiently continuous global sections of the holomorphic line bundle of symmetric tensors of rank $(n,0)$ and $(1-n,0)$. The precise continuity condition must be determined by analyzing the ghost field functional integral.

\section{The Point Bundles}

Substantial use was made of the point bundles in the discussions of coordinate cohomology classes in section \ref{ccc}. Here we briefly recall their most basic properties, which are used in section \ref{ccc}.

The assertion is that,  given an arbitrary point $p$ on a compact Riemann surface $M$, there exists a unique holomorphic line bundle $\zeta_p$ that admits a section $f_p \in \Ga(M, \sO(\zeta_p))$ with the divisor $\vartheta(f_p) = 1 \cdot p$. This is the point bundle $\zeta_p$. Existence is obvious as a result of the general procedure of constructing line bundles from divisors. This general procedure essentially implements the diagram-chasing used in establishing the long exact sequence of cohomology groups associated to the short exact sequence of sheaves:
\be\label{divisor}
0 \to \sO^* \xrightarrow{\,\,i\,\,} \sM^*  \xrightarrow{\,\,\vartheta\,\,}  \sD \to 0 
\ee 
where $\sO^*,\sM^*  $ and $\sD$ are, respectively, the sheaf of germs of nowhere-vanishing holomorphic functions, of germs of not identically vanishing meromorphic functions, and of germs of divisors. Their uniqueness is also obvious: given two such line bundles $\zeta_p$ and $\zeta_p'$, the ratio of their defining sections $f_p/f_p' \in \Ga(M, \sO(\zeta_p \zeta_p'^{-1}))$ is a nowhere vanishing holomorphic function, implying  $\zeta_p \zeta_p'^{-1} =1 \in H^1(M, \sO^*)$. 

By (\ref{chern}) these line bundles have $c(\zeta_p)=1$. 

For $ g \geq 1$, $f_p$ is the unique holomorphic section  of $\zeta_p$, modulo a multiplicative constant. For, were there a second, independent holomorphic section $g_p$, the doublet $(f,g)$ would define a map $(f,g): M \to \C\sP^1$ that is a holomorphic bijection, contradicting the assumption that $M$ has genus $g \geq 1$.  To see this in detail, note first that  $(f(q),g(q)), q \in M$ gives a well-defined point in $\C\sP^1$ because under any change of charts on $M$, the two sections transform by multiplying the same multiplicative factor, hence its image in $\C\sP^1$ is chart-independent. Second, the map is surjective, because given any point $(a, b) \in \C\sP^1$, $b \cdot f - a \cdot g $ is also a nontrivial holomorphic section of $\zeta_p$, and hence must vanish at some point by (\ref{chern}). Third, the map is obviously injective because each section of the form $b \cdot f - a \cdot g $ vanishes exactly at one point, again, by  (\ref{chern}). 

This leads us to the conclusions that for $g \geq 1$, $\zeta_p = \zeta_q$ if and only if $p=q \in M$; and that for $g \geq 1$, the map $p \to \zeta_p$ is a bijection from $M$ to the subset of $H^1(M, \sO^*)$ satisfying $c(\xi)= \dim \Ga(M, \sO(\xi))=1$.
 
The situation is very different for $g=0$. The Riemann-Roch theorem (\ref{RR2}) implies that $\dim \Ga(M, \sO(\xi))=2$ if $c(\xi)=1$. Hence any point bundle on $\C\sP^1$ has a complex two-dimensional space of inequivalent holomorphic sections. This, taken with the arguments given above, implies that on $\C\sP^1$, there exists a unique point bundle: $\zeta_p=\zeta_q (\triangleq \zeta)$ for any $p, q \in \C\sP^1$. This of course is because $f(z)=\frac{z-z_p}{z-z_q}$ is a well-defined meromorphic function on $\C\sP^1$, and it (together with its degenerate limits $f(z)=z-a$ and $f(z)=1/(z-a)$) makes all divisors of the form $1 \cdot p$ equivalent. In the standard coordinate covering of $\C\sP^1$,  $\{U_+, z_+\} \cup \{U_-, z_-\}$ with $z_+ \cdot z_- =1$ on the equator, the point bundle $\zeta$ can be represented by the cocycle $\zeta_{+-} =1/z_-$, and the two independent sections can be chosen as $f_+= \{z_+|_{U_+}, 1|_{U_-}\}$ and $f_-=\{1|_{U_+}, z_-|_{U_-}\}$. 

In fact $\zeta$ is the unique line bundle on $\C\sP^1$ with $c=1$. Any line bundle $\xi$ on $\C\sP^1$ of $c(\xi)=n$ is isomorphic to $\zeta^n$, with $\dim \Ga(M, \sO(\xi))=n+1$. A basis of the sections can be chosen to be $f_+^p \cdot f_-^q$, with $p+q=n$, $p, q \geq 0$.

\section{The Chern Class}\label{TCC}

The small amount of results of the Chern class that we need in this note, in particular (\ref{chern}), are very standard material, and are very well explained in many excellent textbooks. The discussion provided in \cite{Gunning} for example is sufficient for our purpose.  In terms of relevance to this work, the small set of notes \cite{XL} may be the most directly related and the most convenient place to look.

\bibliographystyle{unsrt}

\begin{thebibliography}{99}
\bibitem{BPZ} A. A. Belavin, A. M. Polyakov, and A. B. Zamolodchikov, ``Infinite Conformal Symmetry in Two-Dimensional Quantum Field Theory,'' 
Nucl. Phys. {\bf B 241} (1984) 333-380.


\bibitem{Gunning} R. C. Gunning, ``Lectures On Riemann Surfaces,'' 
(Princeton University Press, 1966).


\bibitem{FK} H. M. Farkas and I. Kra, ``Riemann Surfaces,'' {\it Graduate Texts in Mathematics 71} (Springer, 1992).

\bibitem{Hubbard} J. H. Hubbard, ``Teichm\"{u}ller Theory and Applications to Geometry, Topology, and Dynamics,'' Vol. 1
(Matrix Edition, 2006).

\bibitem{Bers} L. Bers, ``Simultaneous Uniformization,'' Bulletin of the American Mathematical Society 66 (1960) 94-97.

\bibitem{HM}
J. Harris and I. Morrison, (1998), ``Moduli of Curves,'' {\it Graduate Texts in Mathematics 187}  (Springer, 1998).



\bibitem{FriedanShenker} D. Friedan and S. Shenker, ``The Analytic Geometry of Two-Dimensional Conformal Field Theory,'' Nucl. Phys. {\bf B 281} (1987) 509-545.

\bibitem{Vafa} C. Vafa, ``Conformal Theories and Punctured Surfaces,'' Phys. Lett. {\bf B 199} (1987) 195-202.

\bibitem{MooreSeiberg1} G. W. Moore and N. Seiberg, ``Polynomial Equations for Rational Conformal Field Theories,'' Phys. Lett. {\bf B 212} (1988) 451-460.

\bibitem{Segal} G. Segal, ``The definition of conformal field theory, '' preprint, 1988; also in U. Tillmann, ed., {\it Topology, geometry and quantum field theory}, pp. 421-577,  London Math. Soc. Lect. Note Ser., Vol. 308. (Cambridge University Press, 2004).

\bibitem{Witten} E. Witten, ``Quantum Field Theory and the Jones Polynomial,'' Commun. Math. Phys. {\bf  121} (1989) 351-399.

\bibitem{Verlinde} H. Verlinde, ``Conformal Field Theory, Two-Dimensional Quantum Gravity and Quantization of Teichm\"{u}ller Space'' Nucl. Phys. {\bf B 337} (1990) 652-680.




\bibitem{Polchinski}
J. Polchinski,
``String Theory. Vol. 1: An Introduction to the Bosonic String,'' {\it Cambridge Monographs on Mathematical Physics}
(Cambridge University Press, 2007).

\bibitem{XL}
X. Liu,
``Notes On Holomorphic Line Bundles Over Compact Riemann Surfaces, Part I,'' {\it to appear}.

\end{thebibliography}

\end{document}